\documentclass[%
 aps,
 prd,
 reprint,
 amsmath,
 amssymb,
 superscriptaddress
]{revtex4-2}

\usepackage{graphicx}
\usepackage{bm} 
\usepackage{amsfonts,amssymb,amsmath,graphicx,color,bm}
\usepackage[linktocpage=true]{hyperref}
\usepackage{enumitem}
\usepackage{orcidlink}
\usepackage{multirow}
\usepackage{subcaption}
\usepackage{booktabs}
\usepackage{caption}
\usepackage{array}
\usepackage{float}
\usepackage{amssymb}
\usepackage{siunitx}

\hypersetup{
colorlinks=true,
citecolor=red,
linkcolor=blue,
urlcolor=blue,
}

\begin{document}

\title{Corrections to inflationary models induced by non-minimal coupling between scalar field and curvature}

	\author{I. V. Fomin\,\href{https://orcid.org/0000-0003-1527-914X}{\orcidlink{}}}
	\affiliation{Bauman Moscow State Technical University, Moscow, 105005, Russia}
	\author{S. V. Chervon\,\href{https://orcid.org/0000-0001-8898-3694}{\orcidlink{}}}
	\affiliation{Bauman Moscow State Technical University, Moscow, 105005, Russia}
	\affiliation{Ulyanovsk State Pedagogical University, Ulyanovsk, 432071, Russia}
	\author{E. S. Dentsel\,\href{https://orcid.org/0009-0000-0629-5004}{\orcidlink{}}}
	\affiliation{Bauman Moscow State Technical University, Moscow, 105005, Russia}
	\author{B. Mishra\,\href{https://orcid.org/0000-0001-5527-3565}{\orcidlink{}}}
	\affiliation{Department of Mathematics, Birla Institute of Technology and Science-Pilani, Hyderabad Campus, Jawahar Nagar, Kapra Mandal, Medchal District, Telangana 500078, India.}

\date{\today}

\begin{abstract}
In this paper, we consider possible corrections to the characteristics of inflationary models based on a specific parametrization of the non-minimal coupling between the scalar field and curvature. At the inflationary stage, these corrections lead to a deformation of the scalar field potential and a corresponding deviation in the determination of the cosmological perturbation parameters. At the same time, it is shown that the proposed parametrization yields a description of the reheating stage dynamics completely analogous to the case of Einstein gravity with minimal coupling between the scalar field and curvature. For a model-independent analysis of inflationary corrections induced by a non-minimal coupling, a classification of inflationary scenarios based on the expansion in series of the dependence of the tensor-to-scalar ratio on the spectral index of scalar perturbations is considered. It is also shown that this approach allows for the inclusion of well-known inflationary models as special cases.
\end{abstract}


\maketitle

\section{Introduction}

The standard inflationary models, based on the Einstein-Hilbert action with a minimally coupled canonical scalar field, provide the most simple and computationally transparent framework for analysing the primordial spectrum of cosmological perturbations~\cite{Baumann:2014nda,Chervon:2019sey,Martin:2013tda}. Nevertheless, observational constraints imposed by the Planck, Atacama Cosmology Telescope (ACT) and other collaborations on the parameters of cosmological perturbations become increasingly stringent~\cite{Planck:2018vyg,BICEP:2021xfz,AtacamaCosmologyTelescope:2025blo,AtacamaCosmologyTelescope:2025nti,DESI:2024mwx}, which leads to the need to consider more complex inflationary models.
The complexity of inflationary models includes various modifications of the canonical inflationary scalar field (or fields)~\cite{Chervon:2013btx,Paliathanasis:2018vru,Chervon:2019nwq,Fomin:2021snm,Gomez-Valent:2024tdb,Barbosa-Cendejas:2017pbo,
Nautiyal:2018lyq,FerreiraJunior:2023qxi}, modifications of Einstein gravity~\cite{Sotiriou:2008rp,Nojiri:2010wj,DeFelice:2010aj,DeFelice:2011bh,DeFelice:2011zh,DeFelice:2011jm,Clifton:2011jh,
Capozziello:2011et,Nojiri:2017ncd,Ishak:2018his,Han:2019mvj,CANTATA:2021asi,Bahamonde:2021gfp,Odintsov:2023weg,BarrosoVarela:2025yyr,Roy:2026vwx}, and a revision of the standard description of the reheating stage~\cite{DEramo:2017gpl,Maharana:2017fui,Haque:2025uis,Figueroa:2019paj,Odintsov:2025bmp}.

In this context, an important role is played by modified gravity theories implying that the propagation speed of tensor perturbations equals the speed of light in vacuum (as in the case of GR). This criterion follows from recent observations of gravitational waves from the neutron star merger event GW170817~\cite{LIGOScientific:2017ync}, in which the estimated deviation was found to be $|c_g - 1|\leq 5 \times 10^{-16}$.
One of the types of modified gravity satisfying this criterion is scalar-tensor gravity theories~\cite{Fujii:2003pa,Faraoni:2004pi}.
Numerous studies have constructed and analysed inflationary scenarios within scalar-tensor gravity theories, using both exact and approximate solutions to the equations of cosmological dynamics~\cite{Garcia-Bellido:1995wdd,Pozdeeva:2016cja,Belinchon:2016lwr,Bhattacharya:2017uwi,Fomin:2018blx,Fomin:2019yls,
Motohashi:2019tyj,Lopez:2021agu,Amaek:2021cqs}. It is also worth noting that inflationary models based on a quadratic relation between the non-minimal coupling function of a scalar field to curvature and the Hubble parameter $F = (H/\lambda)^{2}$ were studied in~\cite{Fomin:2017sbt,Fomin:2020woj,Fomin:2020caa,Fomin:2022ozv}. In these works, it was shown that the proposed coupling makes inflationary models with arbitrary scalar field potentials compatible with the  Planck observational constraints~\cite{Planck:2018vyg,BICEP:2021xfz}. Nevertheless, the relation $F = (H/\lambda)^{2}$ constitutes a special case of the general power-law parametrisation of the non-minimal coupling between the scalar field and curvature. Thus, it is of interest to examine the impact of the non-minimal coupling of the scalar field to curvature in a more general setting, including the possible non-trivial effects induced by scalar-tensor modifications of Einstein gravity. Also of relevance are the analysis of the ``response'' of cosmological models to the proposed modifications of Einstein gravity and the assessment of the possibility of verifying inflationary models against both Planck and ACT observational constraints simultaneously~\cite{Planck:2018vyg,BICEP:2021xfz,AtacamaCosmologyTelescope:2025blo,AtacamaCosmologyTelescope:2025nti,DESI:2024mwx}.

In the present work we adopt a power-law parametrization of the coupling function expressed through the Hubble parameter $F = (H/\lambda)^{2n}$ which enables us to quantify effects of non-minimal coupling between scalar field and curvature in a controlled manner: the index $n$ plays the role not merely of a phenomenological constant but of a measure characterizing the departure of the potential from the form it would possess in the minimally coupled theory. Simultaneously, the normalization scale $\lambda$ is unambiguously fixed by requiring the equality of the scalar perturbation amplitudes for the minimally and non-minimally coupled cases, guaranteeing the correct observational calibration of the models.

In order to carry out a model-independent analysis of the corrections to cosmological perturbation parameters induced by the proposed modifications of Einstein gravity, we make use of the formalism developed in~\cite{Fomin:2024xzm}, which implies a transition from the analysis of isolated potentials to model-independent analysis of the $r = r(1-n_S)$ relation. This approach allows us to analyse how the non-trivial manifestations of non-minimal coupling during the inflationary stage reshape the landscape of viable cosmological scenarios and establish new benchmarks for detecting deviations from Einstein gravity in observational datasets. Furthermore, this approach has enabled the classification of cosmological inflationary models according to the expansion orders of the $r = r(1-n_S)$ relation, as well as by their phenomenological robustness under the proposed modifications of Einstein gravity. The study of the phenomenological robustness of inflationary models under modifications of Einstein gravity is of particular relevance in view of the continually improving precision of observational data, most notably from the Planck satellite and the ACT, as well as the ongoing revision of the reheating stage description. As observational constraints become increasingly stringent, the ability of a given inflationary scenario to retain its viability under theoretical extensions emerges as an important criterion for model selection. This motivates a systematic analysis of how the proposed modifications affect the spectral parameters of cosmological perturbations and whether the resulting predictions remain consistent with observations across a range of well-motivated models.

The paper is organized as follows. Section~\ref{SEC2} provides the essential background on the construction and analysis of cosmological inflation models with a canonical scalar field within the framework of Einstein gravity.
Section~\ref{SEC3} presents the action for inflationary models based on generalised scalar-tensor gravity. The interpretation of these models in the context of a non-minimal coupling between the scalar field and curvature is considered. The conditions necessary for the cosmological dynamics to match the case of Einstein gravity are also analysed within the framework of a power-law parametrisation of the non-minimal coupling of the scalar field to curvature. Furthermore, a physical interpretation was given to the constant parameter $n$ characterising the power-law parametrisation $F = (H/\lambda)^{2n}$ as a quantitative measure of the potential deformation induced by the non-minimal coupling.
Section~\ref{SEC4} considers the cosmological perturbation parameters corresponding to the proposed power-law parametrisation of the non-minimal coupling. It is shown that the constant parameter $n$ possesses a dual interpretation: on the one hand, it characterizes the potential deformation induced by the non-minimal interaction; on the other hand, it governs the systematic shift in the estimated values of cosmological perturbation parameters relative to the standard GR predictions. A important result of this section is the proof that the consistency relation $n_T = -r/8$ is exactly preserved within the proposed parametrization, which distinguishes the considered class of models as phenomenologically unique.
This section also discusses methods for the model-independent estimation of the number of e-folds, the scalar field excursion, and the constraints associated with the standard description of the reheating stage and beyond.
It is noted that the field equation is invariant under the power-law parametrisation of the non-minimal coupling, and the dynamics of the reheating stage are described identically for both the minimal and non-minimal coupling of the scalar field with curvature. The possible influence of a deviation from instant reheating and of alternative models of dark matter production on the estimation of the e-folds number was also considered.
Section~\ref{SEC6} addresses the deviations in the parameters of inflationary models that arise when comparing the minimal and non-minimal coupling of the scalar field to curvature.  It is shown that the normalisation constant $\lambda$ can be interpreted as a scale fixed by matching the scalar perturbation amplitudes in the minimal and non-minimal frameworks, which ensures a proper calibration of theoretical predictions against observations. The influence of the potential deformation parameter $n$ on the energy scale of inflation and on the uncertainties in the determination of the spectral parameters of cosmological perturbations is also considered.
Section~\ref{SEC7} is devoted to the model-independent reconstruction of the inflationary parameters by the orders of expansion of dependence $r = r(1-n_S)$. We present a series-expansion formalism that does not require specifying the specific scalar field potential and rests only on the universal assumptions of slow-roll dynamics and the exit from inflation. Within the first-order expansion we perform a quantitative assessment of the influence of non-minimal coupling on both scalar and tensor perturbations, demonstrating that this effect must be taken into account when confronting theoretical predictions with observational data. It is concluded that a verification of the first-order models predictions against the Planck and  ACT constraints is feasible for $50\leq\Delta N\leq60$, associated with standard reheating.
Furthermore, inflationary models were analysed at the second order of the expansion of the dependence $r = r(1-n_S)$. The key result here is the demonstration of the possibility of a substantial suppression of the tensor perturbation amplitude while the scalar perturbation parameters correspond to GR case. The same section demonstrates the necessity of extending the number of e-folds to $51<\Delta N<90$ for the second-order models to comply with both the Planck and ACT constraints. A summary of our investigations is presented in Section~\ref{SEC10}. The essential expressions for the computation of inflationary model parameters, formulated in terms of both the slow-roll parameters and the Hubble flow parameters, are presented in Appendix~\ref{AppendixA}.

\section{Inflationary models based on the Einstein gravity}~\label{SEC2}

The cosmological models based on the Einstein gravity and a single scalar field are defined by the action~\cite{Baumann:2014nda,Chervon:2019sey}
\begin{eqnarray}
\label{E}
&&S_{E}=\frac{1}{2}\int d^{4}x\sqrt{-g}\,R_{E} \nonumber\\
&&-\int d^{4}x\sqrt{-g}\left[\frac{1}{2}g^{\mu\nu}\partial_{\mu}\phi_{E}\partial_{\nu}\phi_{E}+V_{E}(\phi)\right],
\end{eqnarray}
in the system of units $8\pi G=M^{-2}_{P}=c=\hbar=1$, where reduced Planck mass is $M_{P}=2.4\times10^{18}$ GeV.

The variation of action (\ref{E}) with respect to the metric and field in a spatially flat Friedmann-Robertson-Walker space-time
\begin{eqnarray}
\label{FRW}
ds^2=-dt^2+a^{2}(t)\left(dx^{2}+dy^{2}+dz^{2}\right),
\end{eqnarray}
leads to the following dynamic equations~\cite{Baumann:2014nda,Chervon:2019sey}
\begin{eqnarray}
\label{DE1}
&&3H^{2}_{E}=\frac{1}{2}\dot{\phi}^{2}_{E}+V_{E}(\phi)\equiv\rho_{\phi},\\
\label{DE2}
&&-3H^{2}_{E}-2\dot{H}_{E}=\frac{1}{2}\dot{\phi}^{2}_{E}-V_{E}(\phi)\equiv p_{\phi},\\
\label{DE3}
&&\ddot{\phi}_{E} + 3H_{E}\dot{\phi}_{E} +V'_{E}= 0,
\end{eqnarray}
where $H=\dot{a}/a$, $\rho_{\phi}$ and $p_{\phi}$ are the energy density and the pressure of a scalar field, also, overdot denotes the derivative with respect to cosmic time, and $V'_{E}=\frac{dV_{E}}{d\phi}$.

From the equations (\ref{DE1})--(\ref{DE3}) only two are independent~\cite{Chervon:2019sey}, and this system can be represented in terms of the cosmic time as follows
\begin{eqnarray}
\label{TIME1}
&&V_{E}=3H^{2}_{E}+\dot{H}_{E},\\
\label{TIME2}
&&X_{E}=\frac{1}{2}\dot{\phi}^{2}_{E}=-\dot{H}_{E},
\end{eqnarray}
where $X_{E}$ is kinetic energy of a scalar field.

An alternative formulation of equations (\ref{TIME1})--(\ref{TIME2}) as the Hamilton-Jacobi ones can be obtained by the relation
\begin{eqnarray}
\label{ISB}
&&\dot{H}_{E}=\frac{dH_{E}}{d\phi_{E}}\dot{\phi}_{E},
\end{eqnarray}
due to the Ivanov--Salopek--Bond approach~\cite{Chervon:2019sey}.

Based on expression (\ref{ISB}), equations (\ref{TIME1})--(\ref{TIME2}) can be written as follows
\begin{eqnarray}
\label{E1mF}
&& V_{E}(\phi)=3H^{2}_{E}-2H'^{2}_{E},\\
\label{E2mF}
&&\dot{\phi}_{E}=-2H'_{E}.
\end{eqnarray}

Also, in order to analyse inflationary models, we will consider slow-roll parameters~\cite{Baumann:2014nda,Chervon:2019sey}
\begin{eqnarray}
\label{epsilonex}
&&\epsilon=-\frac{\dot{H}_{E}}{H^{2}_{E}}=\frac{\dot{\phi}^{2}_{E}}{2H^{2}_{E}}=2\left(\frac{H'_{E}}{H_{E}}\right)^{2},\\
\label{deltanex}
&&\delta=-\frac{\ddot{H}_{E}}{2H_{E}\dot{H}_{E}}=-\frac{\ddot{\phi}_{E}}{H_{E}\dot{\phi}_{E}}=2\frac{H''_{E}}{H_{E}}.
\end{eqnarray}

Thus, the ratio of the kinetic energy of a scalar field to the potential can be written as follows
\begin{eqnarray}
\label{XV}
&&\frac{X_{E}}{V_{E}}=\frac{\epsilon}{3-\epsilon},
\end{eqnarray}
while relation between acceleration of a scalar field and friction term is
\begin{eqnarray}
\label{deltanexREL}
&&\frac{\delta}{3}=-\frac{\ddot{\phi}_{E}}{3H_{E}\dot{\phi}_{E}}.
\end{eqnarray}

Thus, cosmological dynamic equations (\ref{DE1})--(\ref{DE3}) can be written in terms of the slow-roll parameters in the following form
 \begin{eqnarray}
\label{CDE1}
&&V_{E}=H^{2}_{E}\left(3-\epsilon\right),\\
\label{CDE2}
&&X_{E}=\frac{1}{2}\dot{\phi}^{2}_{E}=-\dot{H}_{E}=\epsilon H^{2}_{E},\\
\label{CDE3}
&&-\frac{\ddot{\phi}_{E}}{H_{E}\dot{\phi}_{E}}=\delta,\\
\label{CDE4}
&&V'_{E}=-3H_{E}\dot{\phi}_{E}\left(1-\frac{\delta}{3}\right),
\end{eqnarray}
where equations (\ref{CDE3})--(\ref{CDE4}) correspond to field equation (\ref{DE3}).

The equation of state parameter is
\begin{eqnarray}
\label{EOSGR}
&&w_{E}=\frac{p_{\phi}}{\rho_{\phi}}=\frac{X_{E}-V_{E}}{X_{E}+V_{E}}=-1+\frac{2}{3}\epsilon.
\end{eqnarray}

In the quasi-de Sitter regime of accelerated expansion of the early universe $H\approx constant$ ($\epsilon\ll1$), equations (\ref{XV}) and (\ref{EOSGR}) imply $X/V\ll1$ and $w\approx-1$. Also, condition $|\delta|\ll1$ suggests domination of the friction term $3H\dot{\phi}$ over acceleration of a scalar field $\ddot{\phi}$.
These conditions correspond to a slow-rolling of the scalar field to the minimum of the potential at the inflationary stage.

In this approximation, the slow-roll parameters can be expressed in terms of the potential as~\cite{Baumann:2014nda,Chervon:2019sey}
\begin{eqnarray}
\label{FIRST4}
&&\epsilon\simeq\frac{1}{2}\left(\frac{V'_{E}}{V_{E}}\right)^{2},\\
\label{FIRST5}
&&\delta\simeq\frac{V''_{E}}{V_{E}}-\frac{1}{2}\left(\frac{V'_{E}}{V_{E}}\right)^{2}.
\end{eqnarray}

Also, under slow-roll conditions $\epsilon\ll1$, $|\delta|\ll1$, dynamic equations (\ref{CDE1}) and (\ref{CDE4}) can be written as follows
\begin{eqnarray}
\label{SRCOR2}
&&V_{E}\simeq3H^{2}_{E},\\
\label{SRCOR3}
&&\dot{\phi}_{E}=-2H'_{E}.
\end{eqnarray}

A widely used method for analyzing inflationary models is to determine a specific type of potential $V_{E}=V_{E}(\phi)$ and solution of equations (\ref{SRCOR2})--(\ref{SRCOR3}) for this potential~\cite{Baumann:2014nda,Chervon:2019sey,Martin:2013tda}. A comprehensive survey of inflationary models analysed within the slow-roll approximation can be found in the review~\cite{Martin:2013tda}.

\section{The inflationary models based on the generalized scalar-tensor gravity}\label{SEC3}

For inflationary models based on the generalized scalar-tensor (GST) gravity theory we consider the following action~\cite{DeFelice:2011zh,DeFelice:2011jm,Fomin:2017sbt,Fomin:2020woj,Fomin:2020caa,Fomin:2022ozv}
\begin{eqnarray}
\label{ACTIONGST}
&&S_{GST} = \frac{1}{2\kappa}\int d^4x\sqrt{-g}\,F(\phi) R \nonumber\\
&& - \frac{1}{\kappa}\int d^4x\sqrt{-g}\left[\frac{1}{2}\omega(\phi)g^{\mu\nu}\partial_{\mu}\phi \partial_{\nu} \phi + V(\phi)\right],
\end{eqnarray}
where $\omega(\phi)$ and $F(\phi)$ are differentiable functions of a scalar field.

For the case $\kappa = 1$, the geometric scalar field $\phi$ can be interpreted as a physical field, and the given modification of action (\ref{E}) is treated as accounting for the non-minimal coupling of the scalar field to curvature~\cite{DeFelice:2011zh,DeFelice:2011jm,Fomin:2017sbt,Fomin:2020woj,Fomin:2020caa,Fomin:2022ozv}.

The background dynamic equations in a spatially flat four-dimensional FRW space-time corresponding to the action (\ref{ACTIONGST}) in the chosen system of units are~\cite{DeFelice:2011zh,DeFelice:2011jm,Fomin:2017sbt,Fomin:2020woj,Fomin:2020caa,Fomin:2022ozv}
\begin{gather}
\label{DEQ1}
3FH^{2}+3H\dot{F}-\frac{\omega}{2}\dot{\phi}^{2}-V(\phi)=0, \\
\label{DEQ2}
3FH^{2}+2H\dot{F}+2F\dot{H}+\ddot{F}+\frac{\omega}{2}\dot{\phi}^{2}-V(\phi)=0, \\
\label{DEQ3}
\omega\ddot{\phi} + 3\omega H\dot{\phi} + \frac{1}{2}\dot{\phi}^{2}\omega'+V'-6H^{2}F'-3\dot{H}F'=0,
\end{gather}

Here an overdot represents a derivative with respect to the cosmic time $t$, $H \equiv \dot{a}/a$ denotes the Hubble parameter and $F'\equiv dF/d\phi$.

In addition, we note, that the scalar field equation (\ref{DEQ3}) can be derived from (\ref{DEQ1})--(\ref{DEQ2}). For this reason, equations (\ref{DEQ1})--(\ref{DEQ2}) completely describe the cosmological dynamics at the inflationary stage, and can be represented in terms of the field $\phi$ as follows
\begin{eqnarray}
\label{DEQ4}
&&V(\phi)=3FH^{2}+\frac{5}{2}H\dot{F}+F\dot{H}+\frac{1}{2}\ddot{F},\\
\label{DEQ5}
&&X=\frac{1}{2}\omega(\phi)\dot{\phi}^{2}=\frac{1}{2}H\dot{F}-F\dot{H}-\frac{1}{2}\ddot{F}.
\end{eqnarray}

Nevertheless, it should be noted that the field equation (\ref{DEQ3}) is required for analysing the cosmological dynamics during the reheating stage.

\subsection{Power-law parametrization of the non-miminal coupling of a scalar field and curvature}

Now, let us consider the power-law parametrization of the influence of the non-minimal coupling of the scalar field and curvature in the following form
\begin{eqnarray}
\label{PLP}
&&F(t)=\left(\frac{H(t)}{\lambda}\right)^{2n},~~~-1<n<1,
\end{eqnarray}
where $\lambda>0$ and $n$ are the constant parameters.

Within this parametrization, inflationary models with non-minimal coupling of the scalar field and curvature are reduced to the case of Einstein gravity (minimal coupling) $F=1$ in two cases: for $n=0$ with arbitrary inflationary dynamics $H>0$, and for the pure de Sitter stage $H=\lambda$ with any parameter $-1<n<1$.

After substitution (\ref{PLP}) into (\ref{DEQ4})--(\ref{DEQ5}) and taking into account expressions (\ref{epsilonex})--(\ref{deltanex}) we obtain
\begin{eqnarray}
\label{VPLP}
&&\frac{V}{F} =  H^{2} \left[3 + (2n\delta - 5n - 1)\epsilon + (2n^2 - n)\epsilon^{2}\right],\\
\label{XPLP}
&&\frac{X}{F} = H^{2} \left[(1 - n)\epsilon - (2n^2-n) \epsilon^2 - 2n\delta\epsilon \right].
\end{eqnarray}

In order to compare inflationary models with and without non-minimal coupling, let us consider the following conditions
\begin{eqnarray}
\label{TIME2NC}
&&\dot{\phi}^{2} = -2\dot{H},\\
\label{SRCOR}
&&H=H_{E},~~~\phi=\phi_{E},
\end{eqnarray}
corresponding to the same cosmological dynamics and the same evolution of the scalar field in both cases.

Thus, based on equations (\ref{XPLP}) and (\ref{TIME2NC}), we obtain the following expression for kinetic function
\begin{eqnarray}
\label{KINPLP}
&&\omega = \left( \frac{H}{\lambda} \right)^{2n} \big[\, (1 - n) + (n - 2n^2) \epsilon - 2n\delta \,\big].
\end{eqnarray}

Also, after substitution of expressions (\ref{TIME2}), (\ref{VPLP})--(\ref{KINPLP}) into the field equation (\ref{DEQ3}), we obtain the following result,
\begin{eqnarray}
\label{DELTAPLP}
&&-\frac{\ddot{\phi}}{3H\dot{\phi}}=1+\frac{\frac{1}{2}\dot{\phi}^{2}\dot{\omega}+\dot{V}-6H^{2}\dot{F}-3\dot{H}\dot{F}}
{3\omega\dot{\phi}^{2}H}=\frac{\delta}{3},
\end{eqnarray}
where relation $\omega'=\dot{\omega}/\dot{\phi}$ was used.

Therefore, under conditions (\ref{PLP}) and (\ref{TIME2NC}), field equation (\ref{DEQ3})  is completely equivalent to equation (\ref{DE3}) for the case of the minimal coupling between field and curvature
\begin{eqnarray}
\label{DE3M}
&&\ddot{\phi} + 3H\dot{\phi} +V'_{E}= 0.
\end{eqnarray}

Relation between kinetic energy of a scalar field (\ref{XPLP}) and its potential (\ref{VPLP}) is
\begin{eqnarray}
\label{RELPLP}
&&\frac{X}{V}=\frac{\epsilon
[(1 - n) + (n - 2n^2) \epsilon - 2n\delta]}
{3 + (2n\delta - 5n - 1)\epsilon + (2n^2 - n)\epsilon^{2}}.
\end{eqnarray}

The equation of state parameter can be written on the basis of expressions (\ref{VPLP})--(\ref{XPLP}) as follows
\begin{eqnarray}
\label{EOSPLP}
&& w = \frac{p_{\phi}}{\rho_{\phi}}=\frac{X-V}{X+V} \nonumber\\
&& = \frac{3 + [-2 + 4n(\delta - 1)]\epsilon+ 2n(2n - 1)\epsilon^2}{3(2n\epsilon - 1)}.
\end{eqnarray}

From equations (\ref{DELTAPLP}), (\ref{RELPLP})--(\ref{EOSPLP}) under conditions $\epsilon\ll1$, $|\delta|\ll1$ we obtain following expressions for the relation between kinetic term and potential, relation between acceleration of the field and friction term and equation of state parameter at the inflationary stage
\begin{eqnarray}
\label{RELSRC1}
&&\frac{X}{V}\approx\frac{1}{3}(1-n)\epsilon\ll1,\\
\label{RELSRC2}
&&-\frac{\ddot{\phi}}{3H\dot{\phi}}=\frac{\delta}{3}\ll1,\\
\label{RELSRC3}
&&w\simeq-1,
\end{eqnarray}
where upper limit $n<1$ corresponds to the positive non-zero kinetic term $X>0$.

Therefore, under these conditions, the slow-roll regime holds and the scalar field remains close to a vacuum-like state throughout inflation, similarly to the case of inflationary models based on Einstein gravity.

Nevertheless, the satisfaction of conditions (\ref{SRCOR}) at the inflationary stage is ensured by the specific choice of the non-minimal coupling function, the kinetic function, and the scalar field potential.

In the slow-roll approximation, from (\ref{PLP})--(\ref{XPLP}) and (\ref{SRCOR}), we obtain
\begin{eqnarray}
\label{SRCOR4}
&&F(\phi)\simeq\left(\frac{V_{E}(\phi)}{3\lambda^{2}}\right)^{n},\\
\label{SRCOR5}
&&V(\phi)\simeq\left(3\lambda^{2}\right)^{-n}\left[V_{E}(\phi)\right]^{1+n},\\
\label{SRCOR6}
&&\omega(\phi)\simeq(1-n)F(\phi).
\end{eqnarray}

For $n=0$, one recovers the Einstein gravity limit: $F=1$, $\omega=1$, and $V=V_{E}$. In view of expression (\ref{SRCOR5}), the constant $n$ can therefore be regarded as a deformation parameter characterising the distortion of the scalar field potential at the inflationary stage caused by the non-minimal coupling of the field to curvature.

Using expressions~(\ref{SRCOR4})--(\ref{SRCOR6}), the background parameters for the non-minimally coupled case can be reconstructed for any potential $V_{E}=V_{E}(\phi)$ associated with standard  inflation based on Einstein gravity. In addition, however, the unknown parameter $\lambda$ must be fixed and the influence of the non-minimal coupling of the scalar field to curvature on the other parameters of inflationary models  must be evaluated as well.

\section{Parameters of inflationary models}~\label{SEC4}

Let us now consider the parameters of inflationary models that are directly related to their verification against observational data. We begin with the cosmological perturbation parameters, which manifest themselves at the present stage of the evolution of the universe through the anisotropy and polarisation of the CMB. In the following, we consider the conditions for exit from the inflationary stage of accelerated expansion of the early universe in terms of the slow-roll parameters. Finally, we will analyse the number of e-folds and the field excursion, which are directly related to the estimations of the cosmological  perturbation parameters.

\subsection{Parameters of cosmological perturbations}

Current observational bounds on the cosmological perturbation parameters, derived from measurements of CMB anisotropy and polarisation by Planck, BICEP2/Keck Array, and supplemented by the latest ACT and DESI data, are~\cite{Planck:2018vyg,BICEP:2021xfz,AtacamaCosmologyTelescope:2025blo,AtacamaCosmologyTelescope:2025nti,DESI:2024mwx}
\begin{eqnarray}
\label{PS}
&&A_S=(2.1\pm0.03)\times10^{-9}~({\rm Planck/ACT/DESI}),\\
\label{NSP}
&&n_S=0.9649\pm 0.0042~({\rm Planck/BICEP}),\\
\label{NS}
&&n_S=0.9743\pm0.0034~({\rm ACT/DESI}),\\
\label{RP}
&&r<0.1~({\rm Planck}),\\
\label{R}
&&r<0.036~({\rm Planck/BICEP/Keck Array/ACT}),\\
\label{ALPHAP}
&&\alpha_S=-0.006\pm0.013~({\rm Planck}),\\
\label{ALPHAACT}
&&\alpha_S=0.0062\pm0.0052~({\rm ACT/DESI}),
\end{eqnarray}
where $A_S$ and $n_S$ are the amplitude and spectral index of scalar perturbations, and $r$ is the tensor-to-scalar ratio.

Further, we shall calculate the parameters of cosmological perturbations in order to verify the proposed inflationary models against observational constraints. The method of calculating the parameters of cosmological perturbations in the case of cosmological models based on scalar-tensor gravity was considered, for example, in~\cite{DeFelice:2011zh,DeFelice:2011jm}.

Within this approach, scalar perturbations are considered based on the following second-order action
\begin{equation}
S_{2}=\int dt\, d^{3}x\, a^{3}Q_{S}\left[\dot{{\cal R}}^{2}
-\frac{c_{S}^{2}}{a^{2}}\,(\partial{\cal R})^{2}\right]\,,
\label{eq:az2}
\end{equation}
where ${\cal R}$ is the curvature perturbations, and
\begin{eqnarray}
\label{QS}
&&Q_{S}=\frac{w_1 (4w_1w_3+9w_2^2)}{3w_2^2},\\
\label{CS}
&&c_S^2=\frac{3(2 w_1^{2} w_2H-w_2^2 w_4+4 w_1 \dot{w}_1w_2-2w_1^{2}\dot{w}_2)}{w_1(4w_1w_3+9w_2^2)},\\
\label{w1}
&&w_{1}=F,\\
\label{w2}
&&w_{2}=2HF+\dot{F},\\
\label{w3}
&&w_{3}=-9FH^{2}-9H\dot{F}+\frac{3}{2}(H\dot{F}-2F\dot{H}-\ddot{F}),\\
\label{w4}
&&w_{4}=F.
\end{eqnarray}

The velocity of the scalar perturbations in this case $c_S^2=1$ in the chosen system of units~\cite{Fomin:2017sbt,Fomin:2020woj,Fomin:2020caa,Fomin:2022ozv}, which can be verified by direct substitution (\ref{w1})--(\ref{w4}) into (\ref{CS}).

Also, in the slow-roll regime $\epsilon\ll1$, $|\delta|\ll1$, from (\ref{PLP}), (\ref{QS}) and (\ref{w1})--(\ref{w4}) we obtain
\begin{eqnarray}
\label{QS2}
&&Q_S=\frac{\left(\frac{H}{\lambda}\right)^{2n} \epsilon \! \left((n^{2}+n) \epsilon \! - 2n \delta \! - n + 1\right)}{\left(-1 + n \epsilon \!\right)^{2}} \nonumber \\
&&\simeq \left(\frac{H}{\lambda}\right)^{2n}(1-n)\epsilon. 
\end{eqnarray}

The power spectrum of the scalar perturbations is given by the following expression~\cite{DeFelice:2011zh,DeFelice:2011jm}
\begin{eqnarray}
\label{PR}
&&{\cal P}_{S}(k=aH)=A_{S}=\frac{H^{2}}{8\pi^{2}Q_{S}}\simeq \frac{\lambda^{2n}H^{2(1-n)}}{8\pi^{2}(1-n)\epsilon}.
\end{eqnarray}

At the crossing of Hubble radius ($k=aH$) the expression $d\ln k$ can be written as
$d\ln k=(H+\frac{\dot{H}}{H})dt=H(1-\epsilon)dt\simeq Hdt$, where $k$ is the wave number.

Thus, the scalar spectral tilt for models satisfying relation (\ref{PLP}) is given by the following expression
\begin{eqnarray}
\label{nR}
n_{S}-1 &&\equiv \frac{d\ln{\cal P}_{S}}{d\ln k}\bigg|_{k=aH}
\simeq \frac{\dot{{\cal P}}_{S}}{H{\cal P}_{S}}\bigg|_{k=aH} \nonumber\\
&&= -2(2-n)\epsilon_{\ast}+2\delta_{\ast},
\end{eqnarray}
where $\epsilon_{\ast}$ and $\delta_{\ast}$ are the values of the slow-roll parameters at the crossing of the Hubble radius.

The running of the scalar perturbations is
\begin{eqnarray}
\label{RUNPRP}
\alpha_{S} &&\equiv \frac{dn_{{\rm S}}}{d\ln k}\bigg|_{k=aH}
\simeq \frac{1}{H}\left(\frac{dn_{S}}{dt}\right)\bigg|_{k=aH} \nonumber\\
&&= -\frac{2}{H}\left((2-n)\dot{\epsilon}-\dot{\delta}\right)\bigg|_{k=aH}.
\end{eqnarray}

From definition of the slow-roll parameters (\ref{epsilonex})--(\ref{deltanex}) we can write their derivatives as follows
\begin{eqnarray}
\label{EDER}
&&\dot{\epsilon}=2H\epsilon(\epsilon-\delta),\\
\label{DDER}
&&\dot{\delta}=H(\epsilon\delta-\xi),
\end{eqnarray}
where
\begin{eqnarray}
\label{XI}
&&\xi=\frac{1}{2H^{2}}\frac{d}{dt}\left(\frac{\ddot{H}}{\dot{H}}\right)=4\frac{H'H'''}{H^{2}} \nonumber\\
&&\simeq\frac{V'_{E}V'''_{E}}{V^{2}_{E}}-\frac{3}{2}\frac{V''_{E}}{V_{E}}\left(\frac{V'_{E}}{V_{E}}\right)^{2}+
\frac{3}{4}\left(\frac{V'_{E}}{V_{E}}\right)^{4},
\end{eqnarray}
is the third slow-roll parameter.

Thus, the running of the scalar perturbations (\ref{RUNPRP}) can be written as follows
\begin{eqnarray}
\label{RUNPRP2}
&&\alpha_{S}=(-8 + 4n)\epsilon^{2}_{\ast} + (10 - 4n)\delta_{\ast}\epsilon_{\ast} - 2\xi_{\ast}.
\end{eqnarray}

The second-order action for the tensor perturbations is given by expression~\cite{DeFelice:2011zh,DeFelice:2011jm}
\begin{align}
&S_{T}=\sum_{\lambda_{T}=+, \times}\int dt\, d^{3}x\, a^{3} Q_{T}\left[ \dot{h}_{\lambda_{T}}^{2}
-\frac{c_{T}^2}{a^2} (\partial h_{\lambda_{T}})^2 \right],
\label{ST}
\end{align}
where $\lambda_{T}$ corresponds to the summation over two polarizations, and
\begin{eqnarray}
\label{QT}
&&Q_{T}=\frac{1}{4}\, w_{1}=\frac{1}{4}F, \\
\label{cT}
&&c_{T}^{2}=\frac{w_{4}}{w_{1}}=1.
\end{eqnarray}

For tensor perturbations we obtain the following power spectrum
\begin{eqnarray}
\label{PT}
&&{\cal P}_{T}=\frac{H^{2}}{2\pi^{2}Q_{T}}=\frac{\lambda^{2n}H^{2(1-n)}}{2\pi^{2}}.
\end{eqnarray}

Corresponding spectral index of tensor perturbations can be written as
\begin{equation}
\label{nT}
n_{T}\equiv\frac{d\ln{\cal P}_{{\rm T}}}{d\ln k}\bigg|_{k=aH}\simeq\frac{\dot{{\cal P}}_{{\rm T}}}{H{\cal P}_{{\rm T}}}\bigg|_{k=aH}=
-2(1-n)\epsilon_{\ast}.
\end{equation}

Tensor-to-scalar ratio is
\begin{eqnarray}
\label{RQ}
&&r=\frac{{\cal P}_{{\rm T}}}{{\cal P}_{{\rm S}}}=4\frac{Q_{S}}{Q_{T}}=16(1-n)\epsilon_{\ast}.
\end{eqnarray}

Expressions (\ref{nT})--(\ref{RQ}) yield the consistency relation $n_{T}=-r/8$ completely equivalent to the case of Einstein gravity~\cite{Baumann:2014nda,Chervon:2019sey,Martin:2013tda}. Generic scalar-tensor theories predict a violation of this relation, thereby offering a potential channel for their observational discrimination~\cite{Lin:2016gve}. In contrast, the parametrization $F = (H/\lambda)^{2n}$  ensures that the conventional consistency condition holds exactly. This property singles out a phenomenologically distinct class of models in which the non-minimal coupling does not manifest itself through a broken consistency relation but exclusively through the modification of the cosmological perturbation parameters.

As one can see, the power-law parametrization of the non-minimal coupling (\ref{PLP}) implies a red tilted tensor perturbation spectrum $n_{T}<0$ for any inflationary model and for an arbitrary value of the potential deformation parameter $n$ under condition $r>0$ as for the case of Einstein gravity.

For $n=0$, the expressions for the parameters of cosmological perturbations correspond to the case of Einstein gravity, and can be written as follows
\begin{eqnarray}
\label{PREGR}
&&{\cal P}_{S(E)}=\frac{H^{2}}{8\pi^{2}\epsilon},\\
\label{PRETGR}
&&{\cal P}_{T(E)}=\frac{H^{2}}{2\pi^{2}},\\
\label{RGR}
&&r_{(E)}=16\epsilon_{\ast},\\
\label{NE}
&&n_{S(E)}-1=-4\epsilon_{\ast}+2\delta_{\ast},\\
\label{NTE}
&&n_{T(E)}=-2\epsilon_{\ast}=-\frac{r_{(E)}}{8},\\
\label{RUNE}
&&\alpha_{S}=-8\epsilon^{2}_{\ast} + 10\delta_{\ast}\epsilon_{\ast} - 2\xi_{\ast}.
\end{eqnarray}

Also, based on expressions (\ref{RGR})--(\ref{NE}), we obtain the following inverse relationships between slow-roll parameters and parameters of cosmological perturbations
\begin{eqnarray}
\label{EPSHK}
&&\epsilon_{\ast}=\frac{r_{(E)}}{16},\\
\label{EPSHK2}
&&\delta_{\ast}=\frac{1}{2}\left(-1+n_{S(E)}+\frac{r_{(E)}}{4}\right).
\end{eqnarray}

Moreover, an important aspect of testing inflationary models is the evaluation of the departure of the predicted scalar perturbation spectrum from a Gaussian one~\cite{Baumann:2014nda}. The non-Gaussianity of the scalar perturbation spectrum in the equilateral
configuration for inflationary models based on action~(\ref{ACTIONGST}) is determined by the following nonlinear parameter~\cite{DeFelice:2011zh,DeFelice:2011jm}
\begin{eqnarray}
\label{NGL}
f_{\mathrm{NL}}^{\mathrm{equil}}\simeq\frac{55}{36}\epsilon_{s}+\frac{5}{12}\eta_{s},
\end{eqnarray}
where
\begin{eqnarray}
\label{NGL2}
\epsilon_{s}=\frac{Q_{S}c^{2}_{S}}{F},\qquad \eta_{s}=\frac{\dot{\epsilon}_{s}}{H\epsilon_{s}}.
\end{eqnarray}

Taking into account expressions (\ref{PLP}) and (\ref{QS2}), from (\ref{NGL})--(\ref{NGL2}) we obtain
\begin{eqnarray}
\label{NGL3}
f_{\mathrm{NL}}^{\mathrm{equil}}\simeq\frac{55}{36}(1-n)\epsilon_{\ast}+\frac{5}{6}(\epsilon_{\ast}-\delta_{\ast}).
\end{eqnarray}

Therefore, the slow-roll conditions $\epsilon_{\ast} \ll 1$ and $|\delta_{\ast}| \ll 1$ lead to negligible non-Gaussianity of the scalar perturbation spectrum, $f_{\mathrm{NL}}^{\mathrm{equil}} \ll 1$, for the inflationary models under consideration, which is consistent with the Planck constraints~\cite{Planck:2018vyg}.

\subsection{Cosmological parameters between the crossing of the Hubble radius and exit from inflation}

We now relate the cosmological perturbation parameters to the number of e-folds between Hubble crossing of the perturbations and the end of inflation. The number of e-folds $N=N(t)$ can be considered as the characteristic quantity that provides a direct link between the cosmological dynamics, the physics of the post-horizon-crossing epoch, and the estimation of the cosmological perturbation parameters.
It can be defined as follows~\cite{Baumann:2014nda,Chervon:2019sey}
\begin{eqnarray}
\label{EFOLDS}
&&N(t)=\ln\left(\frac{a(t)}{a_{k}}\right),~~~~\dot{N}=H,
\end{eqnarray}
where $a_{k}$ is the scale factor at a specific cosmic time.

The definition of the e-folds number (\ref{EFOLDS}) leads to the equation
\begin{eqnarray}
\label{C1}
&&\frac{dN}{d\epsilon}=\frac{dN}{dt}\frac{dt}{d\epsilon}=\frac{H}{\dot{\epsilon}},
\end{eqnarray}
under the condition $\epsilon\neq const$.

Thus, using equations (\ref{EDER}) and (\ref{C1}), one finds the following expression for the derivative of the first slow-roll parameter with respect to the number of e-folds
\begin{eqnarray}
\label{C1V}
&&\epsilon'_{N}=\frac{d\epsilon(N)}{dN}=2\epsilon(\epsilon-\delta).
\end{eqnarray}

Taking into account $\epsilon > 0$, the model-independent condition for the exit from the inflationary stage of accelerated expansion in terms of the slow-roll parameters is defined as
\begin{eqnarray}
\label{C2V}
&&\epsilon'_{N}\propto(\epsilon-\delta)>0.
\end{eqnarray}

Also, from (\ref{EDER}) and (\ref{C1}), the expression for the difference in the e-folds number between the Hubble radius crossing and the end of inflation can be defined as follows
\begin{eqnarray}
\label{C2}
&&\Delta N=\frac{1}{2}\int^{\epsilon_{e}}_{\epsilon_{\ast}}\frac{d\epsilon}
{\epsilon(\epsilon-\delta(\epsilon))},
\end{eqnarray}
where $\epsilon_{e}=1$, and $\epsilon_{\ast}$ are the values of the first slow-roll parameter at the end of inflation and at crossing of the Hubble radius.

Special case $\delta=\epsilon$ corresponds to the Hubble parameter $H(t)\propto t^{-1}$, and constant slow-roll parameters $\epsilon=\delta=constant$, which does not correspond to the exit from inflation due to condition (\ref{C2V}).

Equation (\ref{C2}) provides an integral relation connecting the slow-roll parameters to the e-fold number. Consequently, for a given functional dependence $\delta = \delta(\epsilon)$, the parameters of cosmological perturbations can be written as functions of the number of e-folds $\Delta N$.

Furthermore, equations (\ref{SRCOR3}) and (\ref{EDER}) lead to the expression
\begin{eqnarray}
\label{C4A}
&&\frac{d\phi}{d\epsilon}=\frac{\dot{\phi}}{\dot{\epsilon}}=\pm\frac{\sqrt{-2\dot{H}}}{2H\epsilon(\epsilon-\delta)}=
\pm\frac{1}{\sqrt{2\epsilon}(\epsilon-\delta)}.
\end{eqnarray}

Therefore, the scalar field excursion between the end of inflation and the Hubble radius crossing is given by the expression
\begin{eqnarray}
\label{C4}
&&|\Delta\phi|=\frac{1}{\sqrt{2}}\int^{\epsilon_{e}}_{\epsilon_{\ast}}\frac{d\epsilon}{\sqrt{\epsilon}(\epsilon-\delta(\epsilon))},
\end{eqnarray}

The Lyth bound sets a lower limit on the field excursion~\cite{Baumann:2014nda}
\begin{eqnarray}
\label{Lyth}
&&|\Delta\phi|\geq {\mathcal O}\left(1\right)\times\left(\frac{r_{E}}{0.01}\right)^{1/2}\nonumber\\
&&={\mathcal O}\left(1\right)\times\left(\frac{r}{0.01(1-n)}\right)^{1/2}.
\end{eqnarray}

Also, due to the Swampland distance conjecture, the conditions for the consistency of inflationary models with quantum gravity and string theory lead to the following upper bound on the scalar field excursion in the field space~\cite{Scalisi:2018eaz}
\begin{eqnarray}
\label{C6}
&&|\Delta\phi|<-\ln\left(H_{\ast}\right),
\end{eqnarray}
where $H_{\ast}$ is the value of the Hubble parameter at the crossing of the Hubble radius.

\subsection{Reheating constraints on the e-folds number}~\label{SEC5}

After the end of inflation, the slow-roll conditions are violated and the reheating stage begins, which implies oscillations of the scalar field near the minimum of the potential~\cite{Turner:1983he,Martin:2010kz,Martin:2014nya,Garcia:2020wiy}.
In the models under consideration the reheating stage can be considered by the same field equation (\ref{DE3M}) for the non-minimal coupling and minimal coupling. Thus, for the power-law parametrisation of the non-minimal coupling between the scalar field and curvature (\ref{PLP}), the description of the post-inflationary dynamics during reheating is identical to that in Einstein gravity.

During reheating, the energy stored in a coherent scalar field condensate is transferred to ordinary relativistic plasma through parametric resonance and subsequent thermalization. The duration and efficiency of this stage determine the temperature at the onset of the radiation-dominated epoch, the gravitational wave spectrum, and can influence predictions for the relic abundance of dark matter particles, making reheating an essential link between inflationary dynamics and standard Big Bang cosmology~\cite{Turner:1983he,Martin:2010kz,Martin:2014nya,Garcia:2020wiy}. Thus, reconciling any inflationary model with observational constraints requires simultaneously considering characteristics of the scalar and tensor perturbations and a physically sound reheating scenario that does not violate nucleosynthesis and is consistent with CMB data.

Since the reheating stage determines the relationship between the inflationary stage and the BBN stage, BBN-constraints are used to estimate the  e-folds number between the Hubble radius crossing and the end of inflation. Nevertheless, modifications of the pre-BBN universe history, in particular a modified expansion rate of the universe, can significantly alter the estimation of the e-folds number.  The possible realization of a modified cosmological history between reheating and BBN stages involves an additional scalar field $\tilde{\phi}$, which may be produced during reheating stage as one of the decay products of the inflaton, together with the ordinary matter fields~\cite{DEramo:2017gpl,Maharana:2017fui}. The presence of an additional scalar field corresponds to the possibility of an alternative description of dark matter production~\cite{DEramo:2017gpl,Maharana:2017fui}.

The generalized expression for the e-folds number between the crossing of the Hubble radius and the end of inflation was considered in following form~\cite{Maharana:2017fui}
\begin{equation}
\label{REHN}
\Delta N \approx 57-\frac{1}{4}\left(1-3w_{reh}\right)\Delta N_{reh}+\frac{1}{4}\ln r+\frac{1}{4}\ln\eta,
\end{equation}
where $w_{reh}$ is the effective equation of state parameter average value of the state parameter over all field oscillations and $\Delta N_{reh}$ is the e-folds number during reheating stage. The last term in expression (\ref{REHN}) describes the influence of the additional scalar field $\tilde{\phi}$ on the e-folds number, with $\eta$ representing the ratio of the total energy density to the radiation energy density at the end of reheating~\cite{Maharana:2017fui}.

Parameter $\eta$ was defined in~\cite{DEramo:2017gpl,Maharana:2017fui} as follows
\begin{eqnarray}
\label{REHETA}
&&\eta\approx\frac{g(T_{r})}{g(T_{reh})}\left(\frac{g_{s}(T_{reh})}{g_{s}(T_{r})}\right)^{(4+q)/3}\left(\frac{T_{reh}}{T_{r}}\right)^{q},
\end{eqnarray}
where $g(T)$ is the number of degrees of freedom associated with the energy density of the radiation at the reheating temperature $T_{reh}$, $g_{s}(T)$ is the effective number of degrees of freedom associated with the entropy density at the reheating temperature $T_{reh}$, $T_{r}$ is the reference temperature restricted due to the BBN constraints, and parameter $q\geq0$ characterizes the deviations from radiation-like equation of state when considering scenarios for dark matter production.
According to the results given in~\cite{Maharana:2017fui}, the possible maximal correction to the e-folds number inspired by the alternative scenarios for dark matter production can be estimated as $0\leq\Delta N_{\eta}\leq15$ due to the BBN constraints on the reheating temperature $T_{reh}$.

For the standard scenarios with $\eta=1$ expression (\ref{REHN}) is reduced to the following form
\begin{eqnarray}
\label{REHN2}
&&\Delta N \approx 57-\frac{1}{4}\left(1-3w_{reh}\right)\Delta N_{reh}+\frac{1}{4}\ln r.
\end{eqnarray}
and e-folds number between the crossing of the Hubble radius and the end of inflation is estimated as $\Delta N_{s}\simeq 50-60$~\cite{Martin:2010kz}.

Thus, estimation of the e-folds number $\Delta N$ between the crossing of the Hubble radius and the end of inflation depends on the effective equation of state parameter $w_{reh}$, e-folds number during reheating stage $\Delta N_{reh}$ (duration of the reheating stage) and the mechanism of dark matter production.

For the instantaneous reheating expression (\ref{REHN}) is reduced to the following form
\begin{eqnarray}
\label{REHN1}
&&\Delta N \approx 57+\frac{1}{4}\ln r + \frac{1}{4}\ln\eta.
\end{eqnarray}

On the other hand, for $w_{reh}>1/3$ the duration of the reheating stage $\Delta N_{reh}$ influences the estimate of the e-folds number between crossing the Hubble radius and the end of inflation $\Delta N$. The influence of the effective state parameter $w_{reh}$ on the characteristics of inflationary models was considered, for example, in~\cite{Haque:2025uis}. At the same time, we note that the BBN/CMB and LIGO constraints on the characteristics of relic gravitational waves lead to the following restriction on the effective equation of state parameter of a scalar field at the reheating stage $w_{reh}\leq0.56$~\cite{Figueroa:2019paj}.

Further, we will consider the number of e-folds between crossing the Hubble radius and the end of inflation, decomposed as
\begin{eqnarray}
\label{REHCGENN}
\Delta N = \Delta N_{s} + \Delta N_{add},
\end{eqnarray}
where $\Delta N_{s} \simeq 50-60$ corresponds to the standard description of the reheating stage, while $\Delta N_{\rm add}$ accounts for significant deviations from instantaneous reheating and/or implies the inclusion of the influence of an additional scalar field on the cosmological dynamics in alternative dark matter production scenarios.

\section{Deviations in the inflationary parameters}~\label{SEC6}

We now proceed to assess the impact of the non-minimal coupling between the scalar field and curvature on the inflationary energy scale and on the parameters of cosmological perturbations, as well as to define the parameter $\lambda$.

Firstly, we analyse condition (\ref{PS}) on the amplitude of scalar perturbations. In order to determine the constant parameter $\lambda$ let us consider the same amplitude of the power spectrum of scalar perturbations at the crossing of the Hubble radius for the case of non-minimal and minimal coupling of the scalar field and curvature
\begin{eqnarray}
\label{PSEQ}
&&A_{S}={\cal P}_{{\rm S}}(k=aH)= \frac{\lambda^{2n}H^{2(1-n)}_{\ast}}{8\pi^{2}(1-n)\epsilon_{\ast}} \nonumber\\
&&=\frac{H^{2}_{\ast}}{8\pi^{2}\epsilon_{\ast}}\simeq2.1\times10^{-9}.
\end{eqnarray}

Condition (\ref{PSEQ}) is implied by the fact that inflationary models in both the minimal and non-minimal coupling cases can produce the correct amplitude of scalar perturbations (\ref{PS}).

Equation (\ref{PSEQ}) leads to the following relation between the constant parameter $\lambda$ in the power-law parametrization of non-minimal coupling function (\ref{PLP}) and the value of the Hubble parameter at the crossing of the Hubble radius
\begin{eqnarray}
\label{LPLPH}
&&\left(\frac{\lambda}{H_{\ast}}\right)^{2n}=1-n,
\end{eqnarray}
where the value of the Hubble parameter at the crossing of the Hubble radius is defined from (\ref{PSEQ}) as follows
\begin{eqnarray}
\label{HSC}
&&H_{\ast}=\frac{\pi}{2}\sqrt{\frac{2rA_{S}}{1-n}}.
\end{eqnarray}

Further, based on relation
\begin{eqnarray}
\label{HSCV}
&&H\simeq\left(\frac{V_{E}}{3}\right)^{1/2},
\end{eqnarray}
and expression (\ref{LPLPH}) we can exclude the constant $\lambda$ and rewrite equations (\ref{SRCOR4})--(\ref{SRCOR6}) in the following form
\begin{eqnarray}
\label{SRCOR5L3}
&&V(\phi)\simeq\left(\frac{1}{1-n}\right)\frac{\left[V_{E}(\phi)\right]^{n+1}}{\left[V_{E(\ast)}\right]^{n}}
\sim\left[V_{E}(\phi)\right]^{n+1},\\
\label{SRCOR4L3}
&&F(\phi)\simeq\left(\frac{1}{1-n}\right)\left[\frac{V_{E}(\phi)}{V_{E(\ast)}}\right]^{n}
\sim\left[V_{E}(\phi)\right]^{n},\\
\label{SRCOR6L3}
&&\omega(\phi)\simeq\left[\frac{V_{E}(\phi)}{V_{E(\ast)}}\right]^{n}\sim(1-n)\left[V_{E}(\phi)\right]^{n},
\end{eqnarray}
where lower limit $-1<n$ corresponds to the deviation of the potential from a flat one $V\neq const$.

From (\ref{SRCOR6L3}) it follows that the value of the kinetic function at the crossing of the Hubble radius $\omega_{\ast}\simeq1$, and the value of the coupling function is $F_{\ast}\simeq(1-n)^{-1}$.

Also, from (\ref{SRCOR5}), (\ref{LPLPH}) and (\ref{HSCV}) we can estimate the energy scale of inflation as
\begin{eqnarray}
\label{ENSCALE}
&&V^{1/4}_{\ast}\simeq\left(\frac{3H^{2}_{\ast}}{1-n}\right)^{1/4}.
\end{eqnarray}

Further, from (\ref{nR}) and (\ref{NE}), we obtain the following difference in the tilt of the spectrum of scalar perturbations
\begin{eqnarray}
\label{NSD}
&&\Delta n_{S}=n_{S}-n_{S(E)}=2n\epsilon_{\ast}=\frac{n}{8}r_{(E)}.
\end{eqnarray}

Difference in the running of the scalar perturbations can be written on the basis of expression (\ref{RUNPRP2}) as
\begin{eqnarray}
\label{RUNDIFF}
&&\Delta\alpha_{S}=\alpha_{S}-\alpha_{S(E)}=4n\epsilon_{\ast}(\epsilon_{\ast}-\delta_{\ast}) \nonumber\\
&&=\frac{n}{8}r_{(E)}\left(1-n_{S(E)}-\frac{r_{(E)}}{8}\right) \nonumber\\
&&=\Delta n_{S}\left(1-n_{S(E)}+n_{T(E)}\right).
\end{eqnarray}

From expressions (\ref{nT}) and (\ref{NTE}), we can associate the constant parameter $n$ in the power-law parametrization of non-minimal coupling function (\ref{PLP}) with a relative deviation in the estimation of the tilt of the spectrum of tensor perturbations for the case of the Einstein gravity
\begin{eqnarray}
\label{NTD}
&&n=\frac{n_{T(E)}-n_{T}}{n_{T(E)}}.
\end{eqnarray}

On the other hand, expressions (\ref{RQ}) and (\ref{RGR}) imply that we can associate the constant parameter $n$ with the same relative deviation in the estimation of the tensor-to-scalar ratio for the case of the Einstein gravity
\begin{eqnarray}
\label{RTSD}
&&n=\frac{r_{(E)}-r}{r_{(E)}}=\frac{n_{T(E)}-n_{T}}{n_{T(E)}},
\end{eqnarray}
where
\begin{eqnarray}
\label{RNEW}
&&r=(1-n)r_{(E)},\\
\label{NTNEW}
&&n_{T}=(1-n)n_{T(E)}.
\end{eqnarray}

Thus, the constant $n$ on the one hand characterizes the deformations of the scalar field potential (\ref{SRCOR5L3}) induced by non-minimal coupling of a scalar field and curvature, and on the other hand it determines corresponding deviations in the values of the parameters of cosmological perturbations from the case of Einstein gravity.

From expressions (\ref{SRCOR5L3}) and (\ref{RNEW}), it can be concluded that the deviation from the flat (constant) potential $V\neq const$ and the positive value of the tensor-scalar ratio $r>0$ determine the following restrictions on the potential deformation parameter $-1<n<1$.
We also note that inflationary models with power-law parametrization of the non-minimal coupling of a scalar field and curvature (\ref{PLP}) for the special case $n=1$ were considered earlier in~\cite{Fomin:2017sbt,Fomin:2020woj,Fomin:2020caa,Fomin:2022ozv}.

\section{The analysis of inflationary models by dependence $r=r(1-n_{S})$}\label{SEC7}

Now we will consider the dependence of the tensor-to-scalar ratio on the spectral index of scalar perturbations $r=r(1-n_{S})$. Since the values of the spectral index of scalar perturbations imply $1-n_{S}\simeq0.03\ll1$, we can write the dependence $r=r(1-n_{S})$ as follows~\cite{Fomin:2024xzm}
\begin{eqnarray}
\label{RNSEXPANSION}
&&r=\sum^{\infty}_{k=0}\beta_{k}(1-n_{S})^{k}=\beta_{0}+\beta_{1}(1-n_{S}) \nonumber\\
&&+\beta_{2}(1-n_{S})^{2}+...,
\end{eqnarray}
where $(1-n_{S})\ll1$ is the small parameter of expansion and $\beta_{k}$ are the constant coefficients.

Based on this approach, we can estimate model-independent values of the cosmological perturbation parameters for the different orders of the series expansion of the dependence $r = r(1-n_{S})$. We also point out that the proposed classification scheme for inflationary models applies equally to Einstein gravity and to any modified gravity theories. In this context, deviations from Einstein gravity influence the values of the constant parameters $\beta_{k}$.

For a generic analysis of inflationary models at an arbitrary separate expansion order in (\ref{RNSEXPANSION}), it is possible to adopt the following relation between the slow-roll parameters
\begin{eqnarray}
\label{SOm1}
&&\delta-\delta_{0}=-s(\epsilon-\epsilon_{0})^{1/m}+{\cal O}\left(\mbox{higher order terms}\right) \nonumber\\
&&\simeq-s(\epsilon-\epsilon_{0})^{1/m},
\end{eqnarray}
where $s>0$, constants $\epsilon_{0}\sim\epsilon\ll1$, $|\delta_{0}|\sim|\delta|\ll1$, and $m=1,2,3,...$

For $m = 1$, it follows from (\ref{nR}) and (\ref{RQ}) that
\begin{eqnarray}
\label{SOm2}
&&1 - n_{S} = 2(2 - n + s)\epsilon_{\ast} - 2(s\epsilon_{0}+\delta_{0}),\\
\label{SOm3}
&&r\simeq\frac{8(1 - n)(1-n_S +2s\epsilon_0 + 2\delta_0)}{2 - n + s},
\end{eqnarray}
corresponding to expansion (\ref{RNSEXPANSION}) up to first order.

For $m>1$ under condition $|\delta_{\ast}|\gg\epsilon_{\ast}$, from (\ref{nR}) and (\ref{RQ})  we obtain
\begin{eqnarray}
\label{SOm4}
&&1 - n_{S} \simeq-2\delta_{\ast},\\
\label{SOm5}
&&r\simeq16(1-n)\left[\epsilon_{0} + \left(\frac{1 - n_{S}+2\delta_{0}}{2s}\right)^{m}\right],
\end{eqnarray}
where we will consider $\epsilon_{0}=\delta_{0}=0$ for $m>1$ in expansion~(\ref{RNSEXPANSION}) due to neglecting all $(m-1)$--order terms. As we will show below, this choice of constants $\{\epsilon_{0},\delta_{0}\}$ corresponds to physically motivated scalar field potentials.

From (\ref{SOm4}) it follows that, in the case of inflationary models beyond the first order ($m>1$), the non-minimal coupling has a negligible impact on the scalar perturbations. At the same time, expressions (\ref{SOm3}) and (\ref{SOm5}) indicate that the non-minimal coupling between the scalar field and curvature is capable of considerably reducing the expected tensor perturbation amplitude for arbitrary order models.

We now turn to inflationary models at successive orders of the expansion~(\ref{RNSEXPANSION}). We can consider two different approaches to the analysis of inflationary models. The first one relies on the direct computation of the model parameters starting from a given scalar field potential $V = V(\phi)$. The key expressions required for this approach are summarised in Appendix~\ref{AppendixA}. The second strategy is based on a model-independent reconstruction of the inflationary parameter space, taking a functional dependence $\delta = \delta(\epsilon)$ as the starting point. In the following, we elaborate on this second method in more detail. Furthermore, from the possible cases, we single out those that correspond to physically motivated scalar field potentials.

\subsection{First-order inflationary models with $\beta_{0}=0$}~\label{FIRST}

Firstly, we consider dependence (\ref{RNSEXPANSION}) in the first order as follows
\begin{eqnarray}
\label{RNSEXPANSION2}
&&r\sim(1-n_{S}).
\end{eqnarray}

Expressions (\ref{nR}), (\ref{RQ}) and (\ref{RGR})--(\ref{NE}) imply that dependence (\ref{RNSEXPANSION2}) corresponds to the linear relation  between the slow-roll parameters
\begin{eqnarray}
\label{GENDEP}
&&\delta=s\epsilon+{\cal O}\left(\mbox{higher order terms}\right)\simeq s\epsilon,
\end{eqnarray}
which is satisfied in the slow-roll approximation, where
\begin{equation}
\label{HOTERMS}
{\cal O}\left(\mbox{higher order terms}\right)\sim\left\{\epsilon^{2},\delta^{2},...,\epsilon',\delta',...\right\},
\end{equation}
and $s$ is some non-zero constant.

The exit criterion for inflation (\ref{C2V}) in this case takes the following form
\begin{eqnarray}
\label{NGCONDEXIT}
\epsilon(1-s)>0,
\end{eqnarray}
and it can be satisfied under the condition $s<1$ for $\epsilon>0$.

An example of such models is inflation with the power-law potential $V_{E}(\phi) \sim \phi^{p}$~\cite{Martin:2013nzq,Gron:2018rtj,Mishra:2018dtg,Avdeev:2022ilo,Odintsov:2023rqf}, which obeys the relation (\ref{GENDEP}).

For $\delta=s\epsilon$, from (\ref{C2}) and (\ref{C4}) we obtain the following expressions
\begin{eqnarray}
\label{C5}
&&\epsilon_{\ast}=\frac{1}{1+2(1-s)\Delta N},\\
\label{C5D}
&&\delta_{\ast}=\frac{s}{1+2(1-s)\Delta N},\\
\label{C6D}
&&|\Delta\phi|=\frac{2\sqrt{2}\Delta N}{1 + \sqrt{1 + 2(1 - s)\Delta N}},
\end{eqnarray}
which will make it possible to determine the parameters of inflationary models for specified values of $s$ and $\Delta N$.

Further, from expressions (\ref{nR}), (\ref{RQ}) and (\ref{C5})--(\ref{C5D}) we get
\begin{eqnarray}
\label{C7}
&&n_{S}=1-\frac{2(2-s-n)}{1+2(1-s)\Delta N},\\
\label{C8}
&&r=\frac{16(1-n)}{1+2(1-s)\Delta N},\\
\label{C9}
&&r=8\left(\frac{1-n}{2-s-n}\right)(1-n_{S}),
\end{eqnarray}
where $s<1$ and $2-s-n>0$.

For relation $\delta=s\epsilon$ from (\ref{EDER})--(\ref{DDER}) we obtain additional connection between the slow-roll parameters for the first-order models
 \begin{eqnarray}
\label{C10x}
&&\xi=s(2s-1)\epsilon^{2}.
\end{eqnarray}

Thus, expressions (\ref{RUNPRP2}), (\ref{C5})--(\ref{C5D}) and (\ref{C10x}) lead to the following expression for the running of the scalar perturbations
 \begin{eqnarray}
\label{C11x}
&&\alpha_{S}=-4(1-s)(2-s-n)\epsilon^{2}_{\ast}\nonumber\\
&&=-\frac{4(1-s)(2-s-n)}{\left(1 + 2(1-s)\Delta N\right)^2}\sim-10^{-4}.
\end{eqnarray}

We now estimate the possible values of the spectral index of scalar perturbations in inflationary models of this type.
For this purpose, from expressions (\ref{C7})--(\ref{C8}) we obtain the inverse relations for the constant parameters $n$ and $s$ with the inflationary parameters
\begin{eqnarray}
\label{C12n}
&&n=\frac{16+\left(2r+16n_S-16\right)\Delta N-r}{16+\left(2r+16n_S-16\right)\Delta N},\\
\label{C13s}
&&s=\frac{\left(2r+16n_S-16\right)\Delta N+8(n_S+1)+r}{16+\left(2 r+16n_S-16\right)\Delta N}.
\end{eqnarray}

Constraints on the tensor-to-scalar ratio
\begin{eqnarray}
\label{conditionTSR}
&&0<r<0.036,
\end{eqnarray}
conditions on the constant parameters
\begin{eqnarray}
\label{conditions}
&&-1<n<1,~~s<1,~~2-s-n>0,
\end{eqnarray}
and expressions (\ref{C12n})--(\ref{C13s}) lead to the following upper bounds on the spectral index of the scalar perturbations and e-folds number
\begin{eqnarray}
\label{nSBF}
&&n_{S}<1-\frac{r}{8}<0.9955,\\
\label{NB}
&&\Delta N<\left(1-n_{S}-\frac{r}{8}\right)^{-1}.
\end{eqnarray}

Further, based on the estimation of e-folds number for the standard reheating
\begin{eqnarray}
\label{conditionsN}
&&50\leq\Delta N\leq60,
\end{eqnarray}
we finally obtain following constraints on the spectral index of the scalar perturbations for the first-order models
\begin{eqnarray}
\label{nSBnm}
&&0.98-\frac{r}{8}< n_{S}< 1-\frac{r}{8},\\
\label{nSB}
&&n_{S}=0.9794\pm0.0039.
\end{eqnarray}

Hence, the scalar spectral index in such models is largely inconsistent with Planck constraints (\ref{NSP}), and agrees with ACT data (\ref{NS}) only in a very narrow range.
Also, from (\ref{NB}) it follows that for e-folds number $\Delta N>60$ deviations from the Planck and ACT constraints are increasing for these first-order inflationary models. Therefore, we do not regard this class of inflationary models as viable candidates for complying with both these types of observational bounds.

\subsection{First-order inflationary models with $\beta_{0}\neq0$}~\label{FIRSTB}

Now, we consider inflationary models with a generalised linear dependence between the slow-roll parameters
\begin{eqnarray}
\label{NGCOND}
&&\delta=s\epsilon+b+{\cal O}\left(\mbox{higher order terms}\right) \nonumber\\
&&\simeq s\epsilon+b, 
\end{eqnarray}
where $|b|\ll1$.

The exit criterion for inflation (\ref{C2V}) in this case takes the following form
\begin{eqnarray}
\label{NGCONDEXITF}
\epsilon(1-s)-b>0.
\end{eqnarray}

This criterion can be satisfied under the conditions $\epsilon>0$, $s<1$ and $b\leq 0$ without any additional assumptions.

Further, from (\ref{C2}), (\ref{EDER})--(\ref{DDER}), and (\ref{NGCOND}), we obtain the direct and inverse relations between the e-folds number and the slow-roll parameters
\begin{gather}
\label{NGN}
\Delta N = \frac{1}{2b} \ln\left( \frac{\epsilon_{\ast}(b + s - 1)}{\epsilon_{\ast}(s-1)+b}\right),\\
\label{NG}
\epsilon_{\ast}=\frac{b}{(b + s - 1) e^{-2 \Delta N b} - s + 1},\\
\label{NG2}
\delta_{\ast}=s\epsilon_{\ast}+b=\frac{bs}{(b + s - 1) e^{-2 \Delta N b} - s + 1}+b,\\
\label{NG3}
\xi_{\ast}=\epsilon_{\ast}(2\epsilon_{\ast}s^{2}-s\epsilon_{\ast}+2bs+b)\\
=\frac{2 b^2 \left( \frac{1}{2} + \left( s + \frac{1}{2} \right) (b + s - 1)
 e^{-2 \Delta N b} \right)}{\left( (b + s - 1) e^{-2 \Delta N b} - s + 1 \right)^2}.
\end{gather}

In addition, equations (\ref{C4}) and (\ref{NGCOND}) yield the following expression for the scalar field excursion
\begin{eqnarray}
&&|\Delta \phi| = -\frac{\sqrt{2}}{\sqrt{b(s-1)}} \Bigg[\arctan\left( \sqrt{\frac{s-1}{b}} \right) \nonumber \\
&&- \arctan\left( \sqrt{\frac{s-1}{(b+s-1)e^{-2\Delta N b} - s + 1}} \right) \Bigg], 
\end{eqnarray}

\begin{table*}
\centering
\setlength{\tabcolsep}{12pt}
\begin{tabular}{@{}lSSSSSS@{}}
\toprule
\multirow{2}{*}{\textbf{Parameter}} & \multicolumn{2}{c}{\textbf{$n = 0$}} & \multicolumn{2}{c}{\textbf{$n = 0.5$}} & \multicolumn{2}{c}{\textbf{$n = 0.9$}} \\
\cmidrule(lr){2-3} \cmidrule(lr){4-5} \cmidrule(lr){6-7}
& \textbf{Min} & \textbf{Max} & \textbf{Min} & \textbf{Max} & \textbf{Min} & \textbf{Max} \\
\midrule
$b$ & -0.0163 & -0.0001 & -0.0167 & -0.0001 & -0.0171 & -0.0013 \\
$s$ & -3.0000 & -0.4025 & -3.0000 & -0.1883 & -3.0000 & -0.1347 \\
$\epsilon_*$ & 0.0007 & 0.0023 & 0.0006 & 0.0045 & 0.0006 & 0.0059 \\
$\delta_*$ & -0.0183 & -0.0067 & -0.0186 & -0.0044 & -0.0189 & -0.0047 \\
$\xi_*$ & -0.0000 & 0.0001 & -0.0001 & 0.0001 & -0.0001 & 0.0001 \\
$r$ & 0.0107 & 0.0360 & 0.0052 & 0.0360 & 0.0010 & 0.0095 \\
$n_S$ & 0.9607 & 0.9777 & 0.9607 & 0.9777 & 0.9607 & 0.9777 \\
$n_T$ & -0.0045 & -0.0013 & -0.0045 & -0.0006 & -0.0012 & -0.0001 \\
$\alpha_S$ & -0.0005 & -0.0003 & -0.0005 & -0.0002 & -0.0004 & -0.0002 \\
$|\Delta\phi|$ ($M_P$) & 5.9423 & 10.3527 & 5.8994 & 10.8242 & 5.8640 & 10.8595 \\
$V_*^{1/4}$ (\SI{e16}{GeV}) & 1.0255 & 1.3882 & 1.2099 & 1.9632 & 1.7950 & 3.1478 \\
$H_*$ ($10^{-5}M_P$) & 1.0541 & 1.9315 & 1.0376 & 2.7316 & 1.0213 & 3.1408 \\
$\lambda$ ($10^{-5}M_P$) & {---} & {---} & 0.5188 & 1.3658 & 0.2842 & 0.8740 \\
\bottomrule
\end{tabular}
\caption{Inflationary parameters ranges for the first-order models with relation (\ref{NGCOND}) between the slow-roll parameters corresponding to observational constraints (\ref{PS})--(\ref{ALPHAP}) and restrictions on the field excursion (\ref{Lyth})--(\ref{C6}) for $50\leq\Delta N\leq60$.}
\label{TAB1}
\end{table*}

\begin{figure*}
    \centering
    \begin{subfigure}[b]{0.32\textwidth}
        \centering
        \includegraphics[width=\textwidth]{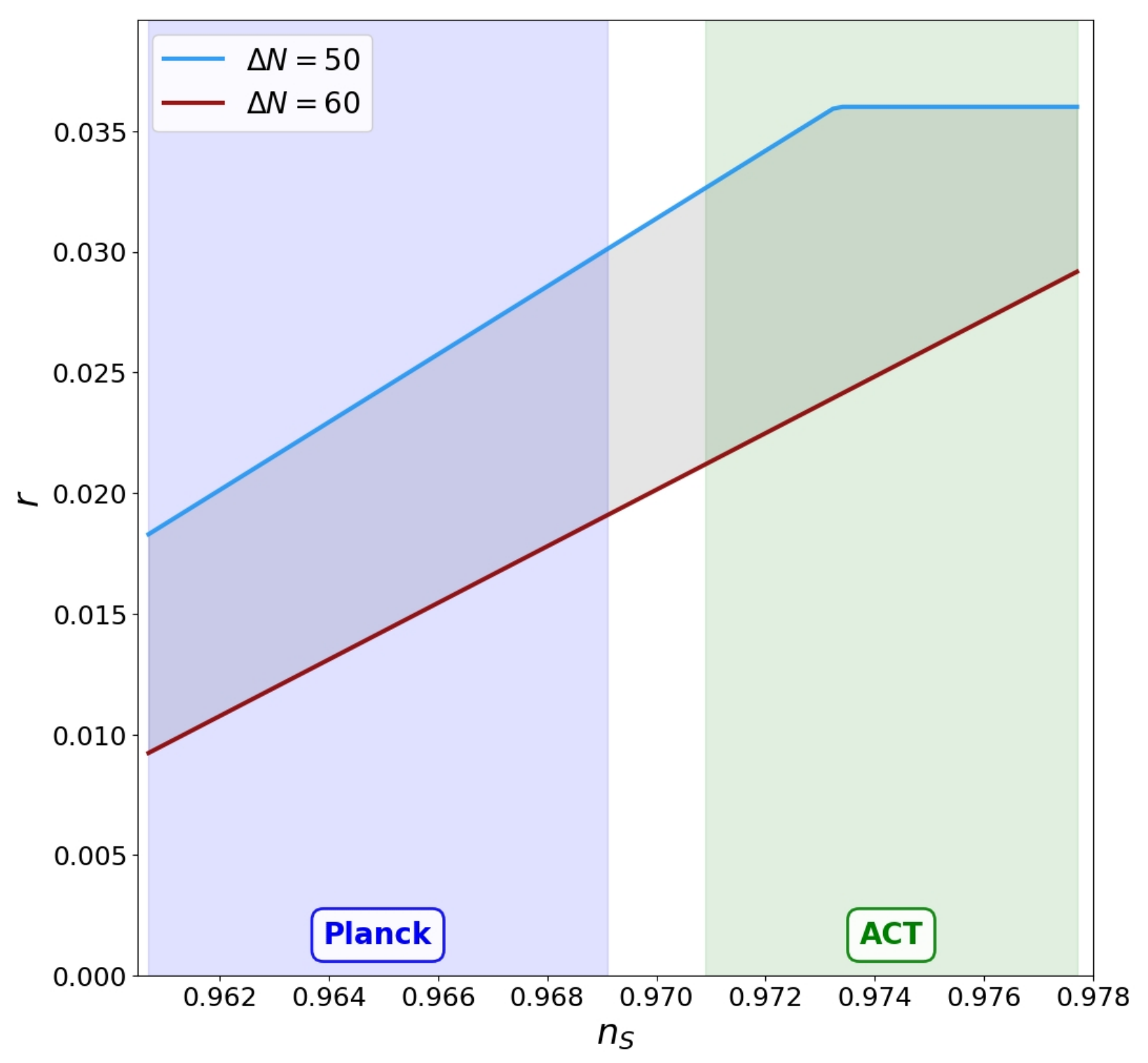}
        \caption{The case $n=0$}
        \label{FIG1}
    \end{subfigure}
    \hfill
    \begin{subfigure}[b]{0.32\textwidth}
        \centering
        \includegraphics[width=\textwidth]{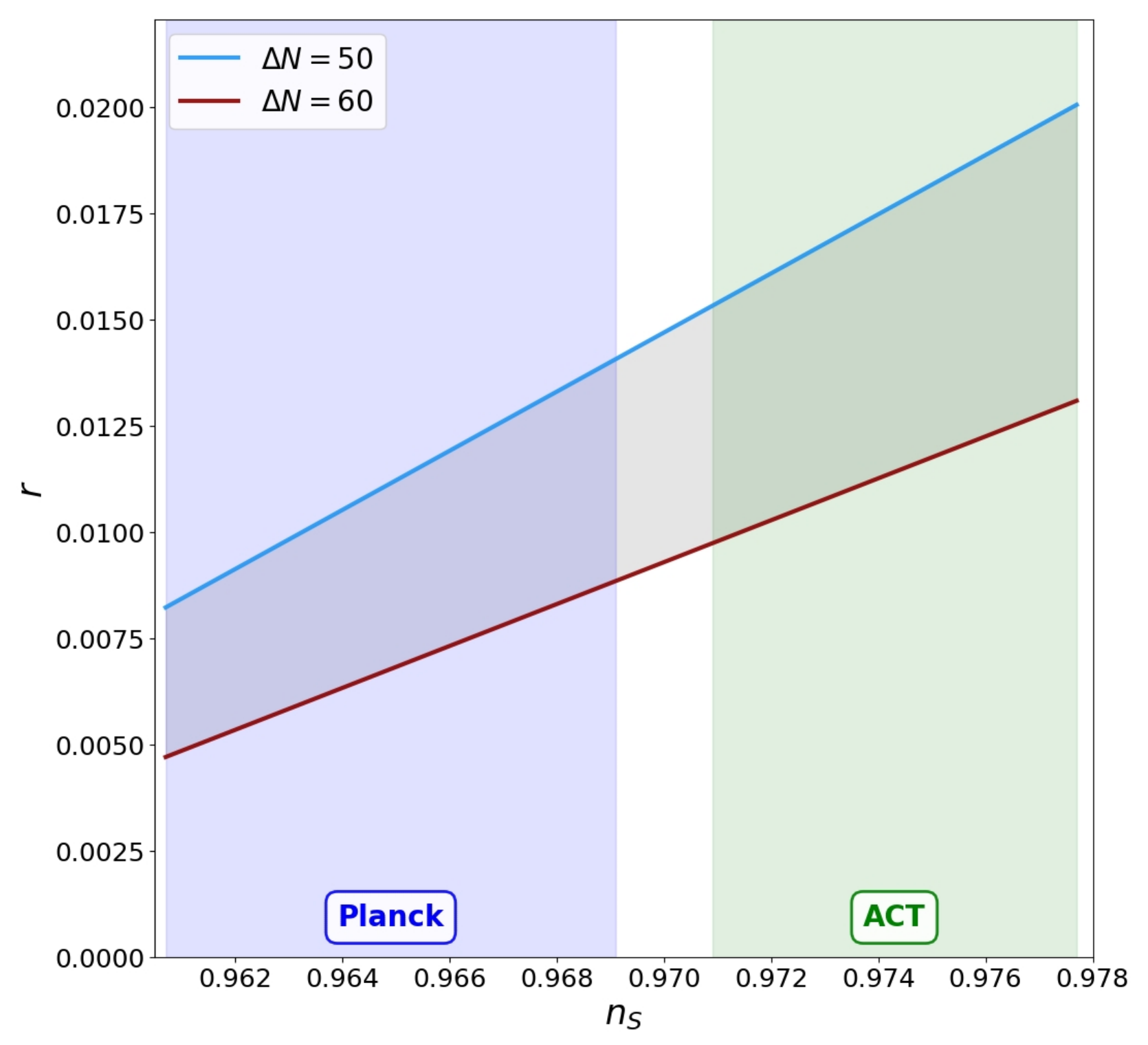}
        \caption{The case $n=0.5$}
        \label{FIG2}
    \end{subfigure}
    \hfill
    \begin{subfigure}[b]{0.32\textwidth}
        \centering
        \includegraphics[width=\textwidth]{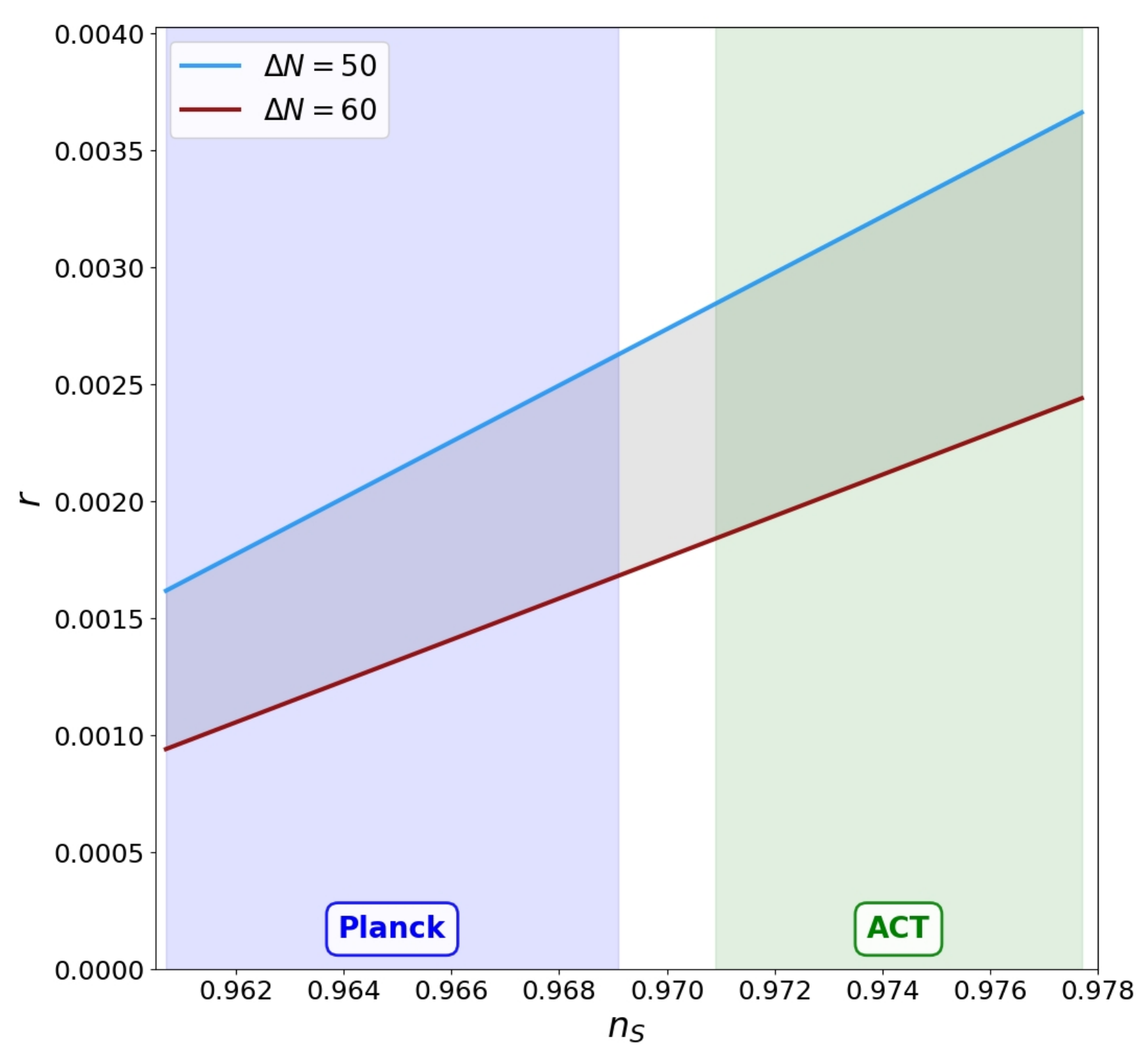}
        \caption{The case $n=0.9$}
        \label{FIG3}
    \end{subfigure}
    \caption{Dependencies $r = r(n_{S})$ for $s=-2.8$ and $n = 0$, $0.5$, $0.9$.
The corresponding scalar field excursion lies in the range $|\Delta\phi/M_{P}|\in[6.1,7.4]$.}
    \label{FIGFIRST}
\end{figure*}

\begin{table*}
\centering
\setlength{\tabcolsep}{8pt}
\begin{tabular}{@{}lcccccc@{}}
\toprule
\multirow{2}{*}{\textbf{Parameter}} & \multicolumn{2}{c}{\textbf{$n = 0$}} & \multicolumn{2}{c}{\textbf{$n = 0.5$}} & \multicolumn{2}{c}{\textbf{$n = 0.9$}} \\
\cmidrule(lr){2-3} \cmidrule(lr){4-5} \cmidrule(lr){6-7}
& \textbf{Planck} & \textbf{ACT} & \textbf{Planck} & \textbf{ACT} & \textbf{Planck} & \textbf{ACT} \\
\midrule
$n_S$ & 0.9626 & 0.9732 & 0.9636 & 0.9751 & 0.9644 & 0.9766 \\
$r$ & 0.0160 & 0.0304 & 0.0080 & 0.0152 & 0.0016 & 0.0030 \\
$\alpha_S$ & $-3.4\times10^{-4}$ & $-4.1\times10^{-4}$ & $-3.1\times10^{-4}$ & $-3.7\times10^{-4}$ & $-2.8\times10^{-4}$ & $-3.3\times10^{-4}$ \\
$a$ & 0.533 & 0.509 & 0.533 & 0.509 & 0.533 & 0.509 \\
$p$ & 0.033 & 0.010 & 0.033 & 0.010 & 0.033 & 0.010 \\
\bottomrule
\end{tabular}
\caption{Parameters for $n = 0, 0.5, 0.9$ with $s=-2.8$, $f=0.9\, M_{P}$ , $\Delta N=55$. Planck: $b=-0.014$, ACT: $b=-0.004$.}
\label{TAB2}
\end{table*}

Furthermore, from expressions (\ref{nR}), (\ref{RQ}), and~(\ref{NGCOND}) there follows a linear relation between the spectral index of scalar perturbations and the tensor-to-scalar ratio
\begin{eqnarray}
\label{NG4}
&&r=\frac{8(1 - n)(1 - n_{S})}{2-n-s}+\frac{16(1 - n)b}{2-n-s}.
\end{eqnarray}

Further, after substituting (\ref{NG})--(\ref{NG3}) into (\ref{nR}), (\ref{RUNPRP2}), and (\ref{RQ}), we obtain the following expressions for the spectral parameters of cosmological perturbations as functions of the e-folds number
\begin{gather}
\label{NG5}
n_{S} = 2\epsilon_{\ast}(n + s - 2) + 2b + 1 \nonumber\\
= \frac{2(b + s - 1)\left(b + \frac{1}{2}\right) e^{-2 \Delta N b}}{(b + s - 1) e^{-2 \Delta N b} - s + 1} \nonumber\\
+\frac{2b(n - 1) - s + 1}{(b + s - 1) e^{-2 \Delta N b} - s + 1}, \\
\label{NG6}
r = 16(1-n)\epsilon_{\ast}=\frac{16b(1 - n)}{(b + s - 1) e^{-2 \Delta N b} - s + 1}, \\
\label{NG8}
\alpha_{S} = -4\epsilon_{\ast}(2-n-s)((1-s)\epsilon_{\ast}-b) \nonumber \\
= -\frac{4b^2 (n + s - 2)(b + s - 1) e^{-2 \Delta N b}}{\bigl( (b + s - 1) e^{-2 \Delta N b} - s + 1 \bigr)^2}. 
\end{gather}

Conditions (\ref{NGCONDEXIT}), $\epsilon_{\ast} > 0$, $s < 1$, $n < 1$, and expression (\ref{NG8}) imply that the running of the scalar perturbations is negative $\alpha_{S} < 0$ independently of the specific choice of parameters in inflationary models with relation (\ref{NGCOND}).

TABLE~\ref{TAB1} presents the allowed ranges of inflationary parameters for first-order models satisfying the slow-roll relation (\ref{NGCOND}), determined by the observational constraints (\ref{PS})--(\ref{ALPHAP}) and the field excursion bound (\ref{C6}) for $50\leq\Delta N\leq60$.
The dependencies $r = r(n_{S})$ for the special case $s = -2.8$ and $n = 0$, $0.5$, $0.9$ are presented in Fig.~\ref{FIGFIRST}.

As can be seen, already for the case of Einstein gravity ($n = 0$), such inflationary models satisfy the observational constraints from both Planck and ACT on the spectral tilt of scalar perturbations and the tensor-to-scalar ratio. For moderate deviations corresponding to $n = 0.5$ and $n = 0.9$, compatibility with the Planck and ACT constraints persists; however, non-minimal coupling between the scalar field and curvature reduces the tensor perturbation amplitude compared to the case of minimal coupling. As a result, the non-minimal coupling between the scalar field and curvature ensures the phenomenological robustness of first-order models against a prospective strengthening of the upper bound on the tensor-to-scalar ratio.

Thus, the first-order models with relation (\ref{NGCOND}) satisfy the observational constraints on the values of the scalar perturbation spectral index and the tensor-to-scalar ratio, with no need to alter the standard treatment of the reheating stage. Nevertheless, the predictions of these models deviate from the ACT constraints on the running of scalar perturbations (\ref{ALPHAACT}) at the level of approximately $1\sigma$.

\subsubsection{Generalized Hybrid Natural Inflation}~\label{GHNISEC}

As the example of inflationary models with relation (\ref{NGCOND}) between the slow-roll parameters we consider the power-law generalized Hybrid Natural Inflation. A notable departure from standard single-field realisations is obtained by combining the inflationary trajectory of Natural Inflation~\cite{Freese:1990rb,Adams:1992bn,Freese:2004un,Kim:2004rp,delaFuente:2014aca} with the exit mechanism characteristic of Hybrid Inflation~\cite{Martin:2013tda}. In this construction, inflation is terminated by a waterfall instability rather than by the steepening of the potential itself. The resulting framework, known as Hybrid Natural Inflation (HNI), was originally introduced and developed through a series of supersymmetric and non-supersymmetric two-field models detailed in~\cite{Ross:2009hg,Stewart:2000pa,Cohn:2000hc,Antusch:2008gw,Ross:2010fg,Vazquez:2014uca}. Owing to the multi-field nature of the scenario, the characteristic scale of the inflaton vacuum expectation value $f$ is not bounded by the Planck scale and can naturally be taken to be super-Planckian~\cite{Martin:2013tda}. A power-law generalisation of hybrid natural inflation was previously considered in~\cite{Munoz:2014eqa,Kitabayashi:2023vfe}.

The scalar field potential for this inflationary model is defined as follows~\cite{Kitabayashi:2023vfe}
\begin{eqnarray}
\label{GHNI}
&&V_{E}(\phi)=V_{0(E)}\left(1+a\cos\left(\frac{\phi}{f}\right)\right)^{p},
\end{eqnarray}
where $0<a<1$ and $p>0$.

For the particular value $p = 1$, the potential coincides with the potential of standard hybrid natural inflation~\cite{Ross:2009hg,Stewart:2000pa,Cohn:2000hc,Antusch:2008gw,Ross:2010fg,Vazquez:2014uca}.

Inflationary slow-roll parameters (\ref{FIRST4})--(\ref{FIRST5}) corresponding to potential (\ref{GHNI}) are
\begin{eqnarray}
\label{GHNI1}
&&\epsilon=\frac{p^2 a^2 \sin^2\left( \dfrac{\phi}{f} \right)}{2f^2 \left( 1 + a \cos\left( \dfrac{\phi}{f} \right)\right)^2},\\
\label{GHNI2}
&&\delta=\frac{a p \bigl( a p \sin^2\left( \dfrac{\phi}{f} \right) - 2\cos\left( \dfrac{\phi}{f} \right) - 2a \bigr)}{2f^2 \bigl( 1 + a \cos\left( \dfrac{\phi}{f} \right) \bigr)^2}.
\end{eqnarray}

Corresponding dependence of the scalar field on the first slow-roll parameter (\ref{GHNI1}) in the slow-roll regime can be written as follows
\begin{eqnarray}
\label{GHNI3}
&&\phi(\epsilon)=\frac{\sqrt{2\epsilon} (a + 1) f^2}{a p}+\mathcal{O}\left(\epsilon^{3/2}\right).
\end{eqnarray}

After substitution (\ref{GHNI3}) into (\ref{GHNI2}) we obtain the following expression for the second slow-roll parameter
\begin{eqnarray}
\label{GHNI4}
&&\delta=\left(\frac{1 + a(p - 2)}{a p}\right)\epsilon-\frac{a p}{(a + 1) f^2}+\mathcal{O}\left(\epsilon^{2}\right)\nonumber\\
&&\simeq s\epsilon+b,
\end{eqnarray}
which coincides with (\ref{NGCOND}) under slow-roll condition $\epsilon\ll1$.

In accordance with expression (\ref{SRCOR5L3}), the modified potential is given by the following expression
\begin{eqnarray}
\label{GHNIM}
&&V(\phi)=A\left(1+a\cos\left(\frac{\phi}{f}\right)\right)^{p(n+1)},
\end{eqnarray}
where
\begin{eqnarray}
\label{POTCONSTANT}
&&A=V_{E(0)}(1-n)^{-1}\left(\frac{V_{E(0)}}{V_{E(\ast)}}\right)^{n}.
\end{eqnarray}

We also note, that under the condition $p(n + 1) = 1$ potential (\ref{GHNIM}) reduces to the form corresponding to standard Hybrid Natural Inflation, but with a non-trivial generalised relation between slow-roll parameters (\ref{GHNI4}).

Equation (\ref{GHNI4}) yields the following expressions for the constant parameters of the model
\begin{eqnarray}
\label{GHNI5}
&& a =  \frac{1 + f^2 (s - 1) b}{2 - f^2 (s - 1) b},\\
\label{GHNI6}
&&p = -\frac{3 b f^2}{1 + f^2 (s - 1) b},
\end{eqnarray}
with the following constraint on the constant parameters
\begin{eqnarray}
\label{GHNI7}
&&\frac{1}{2}<a<1,~~~0 < f < \frac{1}{\sqrt{2(s - 1)b}}\,.
\end{eqnarray}

An example of the evaluation of the cosmological perturbation parameters and the corresponding parameters of Hybrid Natural Inflation is presented in TABLE~\ref{TAB2}. The presented results constitute a special case of the general estimates in TABLE~\ref{TAB1}.

Thus, generalised Hybrid Natural Inflation constitutes a viable model that is consistent with the Planck and ACT observational constraints both in the case of Einstein gravity (minimal coupling) and in the presence of a non-minimal coupling between the scalar field and curvature. Furthermore, the construction of new inflationary scenarios based on the quasi-linear relation (\ref{NGCOND}) between the slow-roll parameters represents a promising direction for future investigation.

\subsection{Second-order inflationary models}\label{SECORD}

Now, we consider conditions $\beta_{0}=0$ and $\beta_{1}=0$ corresponding to the second order in dependence (\ref{RNSEXPANSION}), namely
\begin{eqnarray}
\label{SO}
&&r\sim(1-n_{S})^{2}.
\end{eqnarray}

This dependence corresponds to the  quadratic relationship between slow-roll parameters
\begin{equation}
\label{SO1}
\delta=-s\sqrt{\epsilon}+{\cal O}\left(\mbox{higher order terms}\right)\simeq-s\sqrt{\epsilon},
\end{equation}
which is satisfied in the slow-roll approximation with neglecting the higher order terms, where $s>0$ is some positive non-zero constant.

The condition for exit from inflation
\begin{eqnarray}
\label{SO1EXIT}
\epsilon'_{N} \simeq \epsilon + s\sqrt{\epsilon}>0,
\end{eqnarray}
is satisfied for any $s>0$.

For $\delta\simeq-s\sqrt{\epsilon}$, from (\ref{C2}) and (\ref{C4}) we obtain
\begin{equation}
\label{SO1N}
\Delta N\simeq\frac{1}{s\sqrt{\epsilon_{\ast}}}- \frac{1}{s}-\frac{1}{s^{2}}\ln\left(\frac{(1 + s/\sqrt{\epsilon_{\ast}})}{(1 + s)}\right)
\simeq\frac{1}{s\sqrt{\epsilon_{\ast}}},
\end{equation}
\begin{eqnarray}
\label{SO1P}
&&|\Delta\phi| \simeq \frac{\sqrt{2}}{s} \, \ln\!\left(\frac{1 + s/\sqrt{\epsilon_{\ast}}}{1 + s} \right).
\end{eqnarray}

Therefore, the slow-roll parameters and the scalar field excursion as functions of the e-fold number take the form
\begin{eqnarray}
\label{SO2}
&&\epsilon_{\ast}\simeq\frac{1}{s^{2}(\Delta N)^{2}},\\
\label{SO3}
&&\delta_{\ast}\simeq-\frac{1}{\Delta N},\\
\label{SO4}
&&|\Delta\phi|\simeq\frac{\sqrt{2}}{s}\ln\left(\frac{1+s^{2}\Delta N}{1+s}\right),
\end{eqnarray}
where $s\gg (\Delta N)^{-1}\sim0.01$ due to the slow-roll condition $\epsilon_{\ast}\ll1$.

Further, for relation $\delta=-s\sqrt{\epsilon}$ we obtain
\begin{eqnarray}
\label{SO11}
&&\dot{\delta}=\frac{d\delta}{dt}=\frac{d\delta}{d\epsilon}\dot{\epsilon}=-\frac{s\dot{\epsilon}}{2\sqrt{\epsilon}}.
\end{eqnarray}

Therefore, expressions (\ref{EDER})--(\ref{DDER}) and (\ref{SO11}) lead to the following dependence of the third slow-roll parameter on the e-folds number
\begin{eqnarray}
\label{SO12}
&&\xi_{\ast}=s^{2}\epsilon_{\ast}\simeq\frac{1}{(\Delta N)^{2}}.
\end{eqnarray}

Also, from expressions (\ref{nR}), (\ref{RQ}) and (\ref{SO2})--(\ref{SO3}), we get cosmological perturbations parameters as the functions of the e-folds number
\begin{eqnarray}
\label{SO5}
&&n_{S}\simeq1-\frac{2}{\Delta N}-\frac{2(2-n)}{s^{2}(\Delta N)^{2}}\simeq1-\frac{2}{\Delta N},\\
\label{SO6}
&&r\simeq\frac{16(1-n)}{s^{2}(\Delta N)^{2}},\\
\label{SO7}
&&r\simeq\frac{4}{s^{2}}(1-n)(1-n_{S})^{2}.
\end{eqnarray}

Also, from (\ref{RUNPRP2}), (\ref{SO2})--(\ref{SO3}) and (\ref{SO12}) we obtain running of the scalar perturbations for the second-order models
\begin{eqnarray}
\label{SO13}
&&\alpha_{S}\simeq-\frac{2}{(\Delta N)^{2}}-\frac{2\left(5-2n\right)}{s^{2}(\Delta N)^{3}}-\frac{4(2-n)}{s^{4}(\Delta N)^{4}}\nonumber \\
&&\simeq-\frac{2}{(\Delta N)^{2}}\sim-10^{-4}.
\end{eqnarray}

\begin{table*}
\centering
\setlength{\tabcolsep}{18pt}
\begin{tabular}{@{}lcccc@{}}
\toprule
\multirow{2}{*}{\textbf{Parameter}} & \multicolumn{2}{c}{$s>0$} & \multicolumn{2}{c}{$0<s<2$} \\
\cmidrule(lr){2-3} \cmidrule(lr){4-5}
& \textbf{Min} & \textbf{Max} & \textbf{Min} & \textbf{Max} \\
\midrule
$\Delta N$          & 50.90 & 89.70 & 50.90 & 89.70 \\
$s$                 & 0.236 & 100   & 0.236 & 1.990 \\
$\epsilon_*$        & $\approx0$ & 0.0023 & $\approx0$ & 0.0023 \\
$\delta_*$          & -0.0196 & -0.0112 & -0.0196 & -0.0112 \\
$\xi_*$             & $\approx0$ & 0.0045 & 0.0001 & 0.0045 \\
$r$                 & $\approx0$ & 0.036 & 0.0005 & 0.036 \\
$n_S$               & 0.9607 & 0.9777 & 0.9607 & 0.9777 \\
$n_T$               & -0.0045 & $\approx0$ & -0.0045 & -0.0001 \\
$\alpha_S$          & -0.0008 & -0.0002 & -0.0008 & -0.0002 \\
$|\Delta\phi|$ ($M_P$) & 0.12 & 9.46 & 3.00 & 9.46 \\
$V_*^{1/4}$ (\SI{e16}{GeV}) & 0.067 & 1.39 & 0.48 & 1.39 \\
$H_*$ (\SI{e-5}{M_P}) & 0.0045 & 1.93 & 0.23 & 1.93 \\
\bottomrule
\end{tabular}
\caption{Parameter ranges for second-order model with $n = 0$ and variable $\Delta N$ corresponding to observational constraints (\ref{PS})--(\ref{ALPHAP}) and restrictions on the field excursion (\ref{Lyth})--(\ref{C6}) for $s>0$ and $0<s<2$.}
\label{TAB3}
\end{table*}

\begin{figure*}
    \centering
    \begin{subfigure}[b]{0.32\textwidth}
        \centering
        \includegraphics[width=\textwidth]{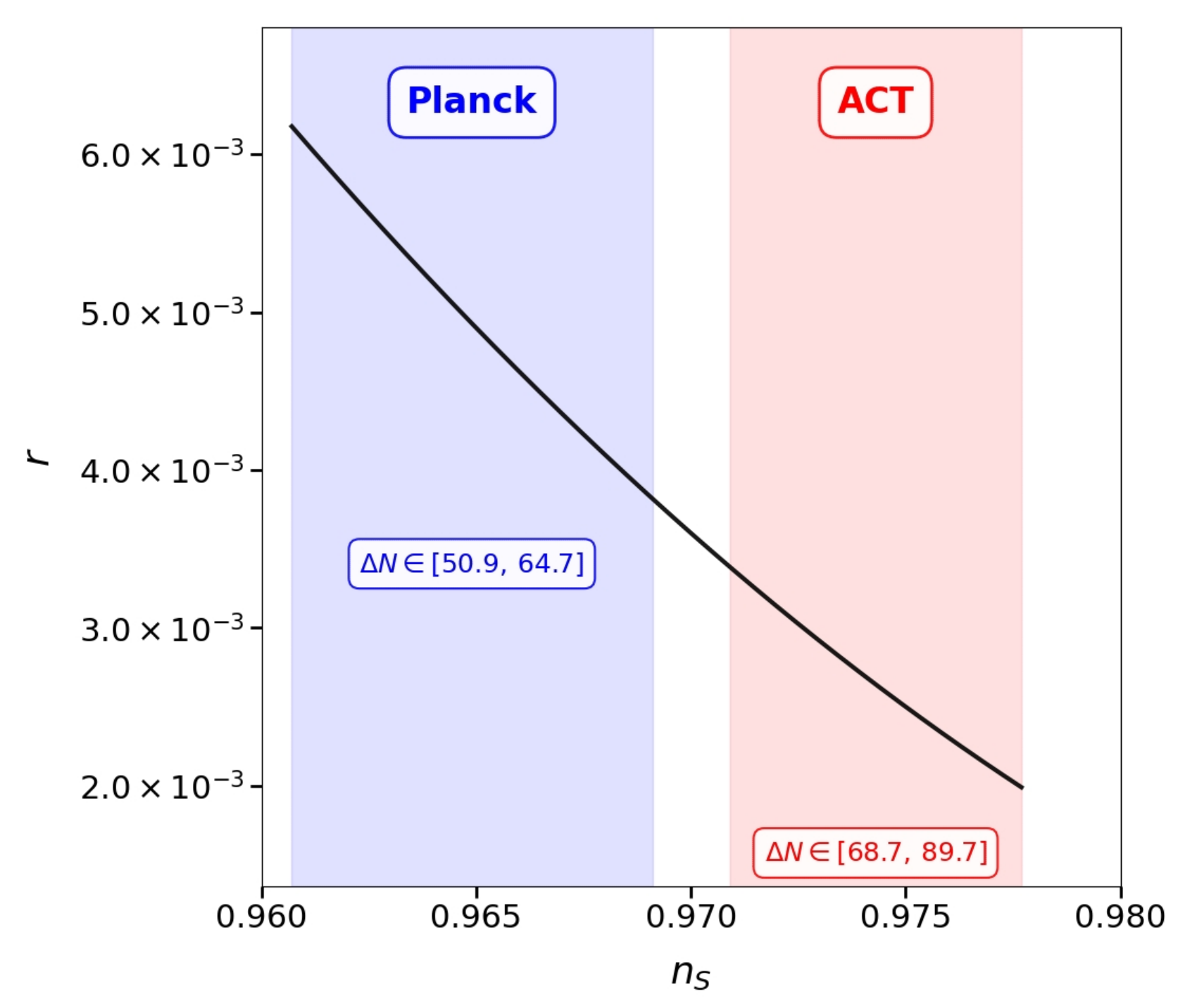}
        \caption{The case $s=1$, $|\Delta\phi/M_{P}|\in[4.6,5.4]$.}
        \label{FIG5}
    \end{subfigure}
    \hfill
    \begin{subfigure}[b]{0.32\textwidth}
        \centering
        \includegraphics[width=\textwidth]{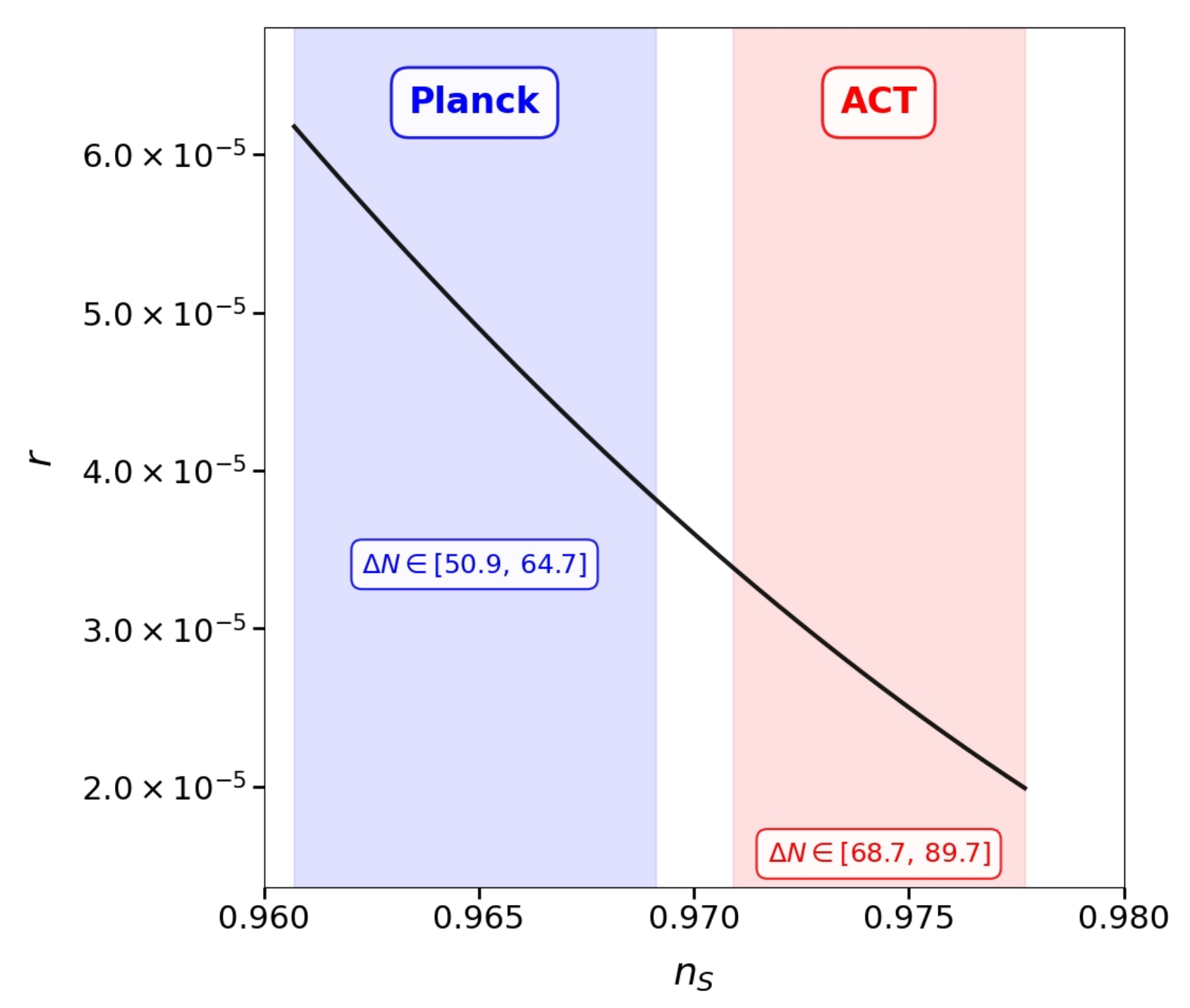}
        \caption{The case $s=10$, $|\Delta\phi/M_{P}|\in[0.87,0.95]$.}
        \label{FIG6}
    \end{subfigure}
    \hfill
    \begin{subfigure}[b]{0.32\textwidth}
        \centering
        \includegraphics[width=\textwidth]{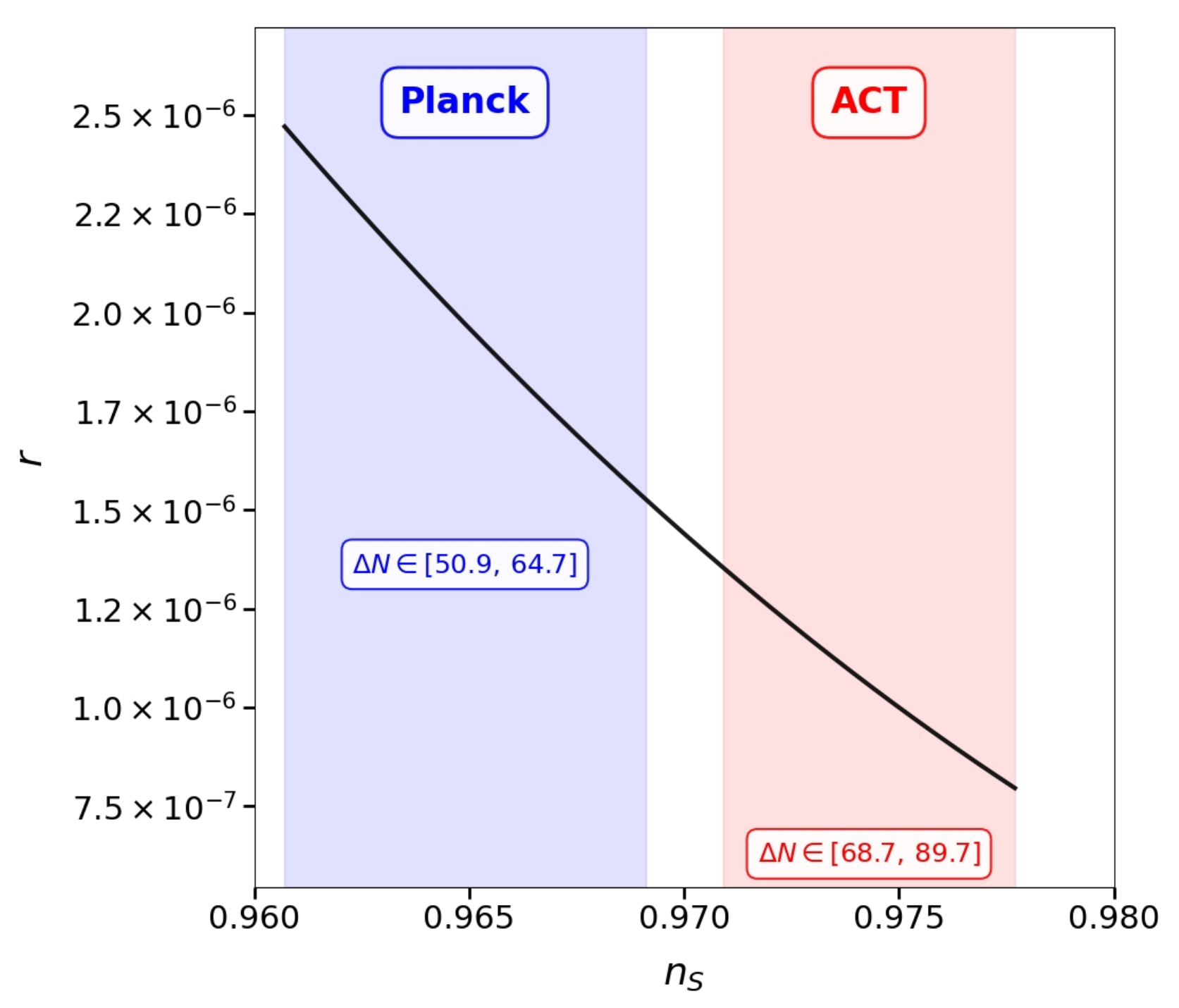}
        \caption{The case $s=50$, $|\Delta\phi/M_{P}|\in[0.22,0.24]$.}
        \label{FIG7}
    \end{subfigure}
    \caption{Dependencies $r = r(n_{S})$ for $s = 1,\,10,\,50$ and $n = 0$ in the second-order models.}
    \label{FIGSEC}
\end{figure*}

Further, from (\ref{NSD})--(\ref{RUNDIFF}) and (\ref{SO5})--(\ref{SO6}), we obtain deviations in the values of the parameters of the scalar perturbations
\begin{eqnarray}
\label{SO8}
&&\Delta n_{S}\simeq\frac{2n}{s^{2}(\Delta N)^{2}}\ll n_{S},\\
\label{SO9}
&&\Delta\alpha_{S}\simeq\frac{4n}{s^{2}(\Delta N)^{3}}+\frac{4n}{s^{4}(\Delta N)^{4}} \ll |\alpha_{S}|.
\end{eqnarray}

Therefore, non-minimal coupling corrections to the characteristics of scalar perturbations in second-order models can be regarded as negligible ones for the second-order models. Nevertheless, expression (\ref{SO6}) indicates that accounting for the non-minimal coupling leads to a possible reduction of the predicted tensor perturbation amplitude in comparison with the minimal coupling scenario $n = 0$.

TABLE~\ref{TAB3} presents the allowed ranges of inflationary parameters for second-order models obeying (\ref{SO1}), as determined by the observational constraints (\ref{PS})--(\ref{ALPHAP}) and the bound on the scalar field excursion (\ref{C6}), with the number of e-folds $\Delta N$ left unrestricted. The dependencies $r = r(n_{S})$ for the special cases $s = 1,\,10,\,50$ and $n = 0$ are presented in Fig.~\ref{FIGSEC}.

A distinctive feature of the second-order models, as compared to the first-order ones, is the possibility of realising inflationary scenarios with a sub-Planckian scalar field excursion $|\Delta\phi|<1$.
Also, it can be noted that the second-order models satisfy the Planck data for $51<\Delta N<65$, and the ACT data for $69<\Delta N<90$. Thus, cosmological models of this type can be consistent with the ACT data only under a revision of the standard description of the reheating stage.

\subsubsection{Modified $\alpha$-attractor models}~\label{EATTR}

As an example of the second-order inflationary model, we consider $\alpha$-attractors models. Cosmological  $\alpha$-attractors models are a class of inflationary theories that arise naturally in the context of supergravity theories. Their distinctive feature is the presence of universal behavior in their predictions, independent of the details of a particular model~\cite{Kallosh:2013yoa,Kallosh:2013hoa,Kallosh:2014rga,Galante:2014ifa,Carrasco:2015rva,German:2021tqs}.

The generalized form of the scalar field potential for $\alpha$-attractor model is defined as follows ~\cite{German:2021tqs}
\begin{eqnarray}
\label{EINF}
&&V_{E}(\phi)=V_{0(E)}\left(1-e^{-k_{1}\phi}\right)^{p},
\end{eqnarray}
where $k_{1}$ and $p$ are some positive constants.

Inflationary slow-roll parameters (\ref{FIRST4})--(\ref{FIRST5}) corresponding to potential (\ref{EINF}) are
\begin{eqnarray}
\label{EINF2}
&&\epsilon=\frac{p^{2} k^{2}_{1} e^{-2k_{1} \phi}}{2 (1 - e^{-k_{1} \phi})^{2}},\\
\label{EINF3}
&&\delta=\frac{k^{2}_{1} p (p - 2 e^{k_{1} \phi})}{2 (e^{k_{1} \phi} - 1)^{2}}.
\end{eqnarray}

The dependence of the scalar field on the first slow-roll parameter $\phi = \phi(\epsilon)$ can be found from (\ref{EINF2}) as
\begin{eqnarray}
\label{EINF4}
&&\phi(\epsilon)=-\frac{1}{k_{1}}\ln\left(\pm\frac{pk_{1}\sqrt{2 \epsilon} + 2 \epsilon}{ 2 \epsilon-p^{2} k^{2}_{1}}\right).
\end{eqnarray}

For the canonical scalar field, and taking into account condition $\epsilon\ll1$, expression (\ref{EINF4}) can be written as follows
\begin{eqnarray}
\label{EINF5}
&&\phi(\epsilon)\simeq-\frac{1}{k_{1}}\ln\left(\frac{\sqrt{2 \epsilon}}{pk_{1}}\right).
\end{eqnarray}

After substitution (\ref{EINF4}) or (\ref{EINF5}) into (\ref{EINF3}) under slow-roll condition $\epsilon\ll1$ we obtain following dependence between slow-roll parameters
\begin{eqnarray}
\label{EINF6}
&&\delta=-k_{1}\sqrt{2\epsilon}+{\mathcal O}\left(\epsilon\right)\simeq-k_{1}\sqrt{2\epsilon},
\end{eqnarray}
corresponding to (\ref{SO1}), where
\begin{eqnarray}
\label{EINF7}
&&k_{1}=\frac{s}{\sqrt{2}}=\sqrt{\frac{2}{3\alpha}}.
\end{eqnarray}

The partial case $m=2$, $p=2$, and $\alpha=1$ ($s=2/\sqrt{3}$) correspond to the Starobinsky inflation~\cite{Kallosh:2013yoa,Kallosh:2013hoa,Kallosh:2014rga,Galante:2014ifa,Carrasco:2015rva,German:2021tqs}.

Further, from expressions (\ref{EINF}) and (\ref{SRCOR5L3}) at the inflationary stage we obtain following modified potential
\begin{equation}
\label{EINF8}
V(\phi)\simeq \left(\frac{V_{0(E)}}{1-n}\right)\left(\frac{V_{0(E)}}{V_{E(\ast)}}\right)^{n}\left(1-e^{-k_{1}\phi}\right)^{p(n+1)},
\end{equation}

Corresponding coupling function $F=F(\phi)$ and kinetic functions $\omega=\omega(\phi)$ at the inflationary stage are defined by expressions (\ref{SRCOR4L3})--(\ref{SRCOR6L3}) with potential (\ref{EINF}).

In addition, we consider the T-attractor model with the potential~\cite{Kallosh:2013yoa,Kallosh:2013hoa,Kallosh:2014rga,Galante:2014ifa,Carrasco:2015rva,German:2021tqs}
\begin{eqnarray}
\label{T1}
&&V_{E}(\phi)=V_{0}\left[\tanh\left(\frac{\phi}{k_{2}}\right)\right]^{p},
\end{eqnarray}
where $k_{2}$ and $p$ are a some positive constants.

Inflationary slow-roll parameters (\ref{FIRST4})--(\ref{FIRST5}) corresponding to potential (\ref{EINF}) are
\begin{eqnarray}
\label{T2}
&&\epsilon=\frac{p^{2}}{2 k^{2}_{2}}{\rm sech}^{2}\left( \frac{\phi}{k_{2}} \right)
{\rm csch}^{2}\left( \frac{\phi}{k_{2}} \right),\\
\label{T3}
&&\delta=C(\phi)
\left(p + 2 -4 \cosh^{2}\left( \frac{\phi}{k_{2}} \right) \right),
\end{eqnarray}
where \(C(\phi)={\rm sech}^{2}\left( \frac{\phi}{k_{2}} \right)
{\rm csch}^{2}\left( \frac{\phi}{k_{2}} \right)\frac{p}{2 k^{2}_{2}}\). 

The dependence of the scalar field from the first slow-roll parameter $\phi=\phi(\epsilon)$ is
\begin{equation}
\label{T4}
\phi(\epsilon)=\frac{k_{2}}{4} \ln\!\left( \frac{\epsilon k^{2}_{2} + 4 p^{2}
\pm 2 \sqrt{2 k^{2}_{2} p^{2}\epsilon
+ 4 p^{4}}}{\epsilon k^{2}_{2}} \right).
\end{equation}

Under condition $\epsilon\ll1$ expression (\ref{T4}) is reduced to the following form
\begin{eqnarray}
\label{T4A}
&&\phi(\epsilon)\simeq \frac{k_{2}}{4}
\ln\!\left( \frac{8p^{2}}{\epsilon k^{2}_{2}}\right),
\end{eqnarray}
for the positive sign.

After substitution (\ref{T4}) or (\ref{T4A}) into (\ref{T3}) under slow-roll condition $\epsilon\ll1$ we obtain following relation between slow-roll parameters
\begin{eqnarray}
\label{T5}
&&\delta=-\frac{2\sqrt{2}}{k_{2}}\sqrt{\epsilon}+{\mathcal O}\left(\epsilon\right)\simeq-\frac{2\sqrt{2}}{k_{2}}\sqrt{\epsilon}.
\end{eqnarray}

Relation (\ref{T5}) corresponds to (\ref{SO1}) for
\begin{eqnarray}
\label{T6K}
&&k_{2}=\frac{2\sqrt{2}}{s}=\sqrt{6\alpha}.
\end{eqnarray}

From expressions (\ref{EINF}) and (\ref{T1}) at the inflationary stage, we obtain following modified potential
\begin{equation}
\label{T6}
V(\phi)\simeq \left(\frac{V_{0(E)}}{1-n}\right)\left(\frac{V_{0(E)}}{V_{E(\ast)}}\right)^{n}\left[\tanh\left(\frac{\phi}{k_{2}}\right)\right]^{p(n+1)},
\end{equation}
where coupling function $F=F(\phi)$ and kinetic functions $\omega=\omega(\phi)$ are defined by expressions (\ref{SRCOR4L3})--(\ref{SRCOR6L3}).

The perturbation parameters for these models follow from (\ref{SO5}), (\ref{SO6}), and (\ref{SO13}) with $s = \frac{2}{\sqrt{3\alpha}}$.
Consequently, the observational verification of these models by ACT necessitates a revised estimate of the number of e-folds between the horizon crossing of perturbations and the end of the inflationary stage. Other examples of second-order inflationary models can be found, for instance, in the review~\cite{Martin:2013tda}.

\subsection{Higher-order inflationary models}

The analysis of higher-order inflationary models based on the ansatz (\ref{SOm1}) faces the technical difficulty that equations (\ref{C2}) and (\ref{C4}) do not admit exact analytical solutions for the e-fold number and the scalar field excursion. However, as previously pointed out, the non-minimal coupling of the scalar field to curvature leaves the scalar perturbation parameters unchanged in this regime, whereas the modified tensor perturbation characteristics can be derived from (\ref{RNEW}) and (\ref{NTNEW}). Higher-order inflationary scenarios, including those motivated by the Minimal Supersymmetric Standard Model or Renormalizable Inflection Point Inflation, were reviewed in~\cite{Martin:2013tda}. Since the predicted tensor-to-scalar ratio in these scenarios lies well below the observational bound (\ref{R}), the inclusion of the non-minimal coupling does not lead to observationally distinguishable predictions, and the models remain phenomenologically equivalent to their minimally coupled counterparts.

\section{Conclusion}~\label{SEC10}

In this paper, we considered the possibility of interpreting modern observational data taking into account the non-minimal coupling between the scalar field and curvature at the inflationary stage of the evolution of the universe. The specificity of the proposed approach was the power-law parametrization of the non-minimal coupling function expressed in terms of the Hubble parameter $F=\left(\frac{H}{\lambda}\right)^{2n}$. In order to compare the predictions of inflationary models based on minimal and non-minimal coupling, the same cosmological dynamics and scalar field evolution were considered for both cases. An additional property of the considered power-law parametrization of the non-minimal coupling is that the consistency relation $n_{T}=-r/8$ is completely equivalent to the case of Einstein gravity. Within the framework of the considered parametrization, the value of the normalization parameter $\lambda$ follows from the equality of the amplitude of scalar perturbations for the case of minimal and non-minimal coupling. The second parameter $n$ has a dual meaning: on the one hand, it characterizes the deformation of the scalar field potential at the inflationary stage induced by non-minimal coupling of a scalar field with the curvature; on the other hand, parameter $n$ determines errors in estimating the parameters of cosmological perturbations for inflationary models based on GR.

The following part of the proposed analysis was a model-independent estimate of the cosmological perturbation parameter values. For this purpose, we used a series expansion of the tensor-scalar ratio dependence on spectral index of scalar perturbations $r=r(1-n_{S})$. Based on this approach, estimates of inflationary parameters were obtained for the entire spectrum of possible scenarios, regardless of the specific potentials of the scalar field at the first and second order of the expansion of dependence $r=r(1-n_{S})$. The only necessary conditions in this case were the implementation of the slow-roll regime at the inflationary stage and the possibility of the exit from inflation.
It was shown that the non-minimal coupling of the scalar field and curvature for inflationary models in the first order of the expansion in a series of dependence $r=r(1-n_{S})$ has a significant effect on both scalar and tensor perturbations. Taking this influence into account, first-order models can be verified against the Planck and ACT constraints for the standard estimate of the e-folds number of between crossing the Hubble radius and the end of inflation.
For second-order models, non-minimal coupling does not have a significant effect on the spectral index of scalar perturbations $n_{S}$, but can significantly reduce the amplitude of tensor perturbations. Also, for the case of a larger number of e-folds $69<\Delta N<90$ than the standard estimates $50\leq\Delta N\leq60$, second-order models also meet the ACT constraints on the spectral index of the scalar perturbations.
The increase in the number of e-folds in this case was considered taking into account the influence of the deviation from instantaneous reheating and the impact of the possible alternative scenarios for dark matter production on the dynamics of the expansion of the universe. It was further shown that the proposed classification scheme implies correspondence with well-known models of cosmological inflation. Moreover, it was noted that higher-order inflationary models imply no phenomenological difference between the cases with and without the non-minimal coupling of the scalar field to curvature.

The properties of the proposed models were assessed based on two criteria: their phenomenological robustness under the proposed modifications of Einstein gravity and the refinement of observational constraints. This approach has made it possible to demonstrate explicitly that early-universe cosmological models at different orders of the expansion $r = r(1-n_{S})$ can be categorized based on the description of the modifications of inflationary stage and the reheating stage.

Nevertheless, it should be noted that in all inflationary models under consideration, running of the scalar spectral index takes the negative values $\alpha_{S}\sim-10^{-4}<0$, whereas the ACT data imply its positive value $\alpha_S=0.0062\pm0.0052>0$. This result is consistent with the predictions of many different inflationary models, both based on Einstein gravity~\cite{Martin:2013tda,Aoki:2025ywt} and extended theories of gravity~\cite{Frolovsky:2025iao,Ketov:2025cqg,Wolf:2025ecy} as well.
Consequently, negative values of the running of the scalar spectral index can be regarded as the primary factor behind the tension ($\sim1\sigma$) between the considered inflationary models and the updated ACT observational constraints.
It should be noted that the recent work~\cite{Roy:2026vwx} also considered inflationary models with a non-minimal coupling between the scalar field and curvature, along with loop quantum corrections. In that work, it was shown that such models allow a running of $\alpha_{S}=0.001$, but with deviations $(\leq 2\sigma)$ of the spectral index of scalar perturbations from the ACT constraints.
Thus, a promising direction for the future development of the proposed approach consists in extending the method considered in this paper to alternative modifications of Einstein gravity in order to incorporate the ACT constraints on the running of scalar perturbations into generalised GR-like inflationary models.

\section*{Acknowledgements}

The work of I.V.F. and S.V.C. was carried out within the framework of Supplementary Agreement No. 073-03-2026-035/1 dated 21.02.2026 to the Agreement on the provision of a subsidy to a federal budgetary or autonomous institution for financial support for the implementation of a state assignment for the provision of public services (performance of work) No. 073-03-2026-035 dated 23.01.2026, concluded between the Federal State Budgetary Educational Institution of Higher Education ``UlGPU named after I.N. Ulyanov" and the Ministry of Education of the Russian Federation.

\appendix
\section{Key expressions of the inflationary parameters}
\label{AppendixA}

In the present appendix, we collect the key formulas for the determination of the background parameters, the slow-roll parameters, and the cosmological perturbation parameters within the framework of a generalised inflationary analysis that accounts for the power-law parametrisation of the effect of the non-minimal coupling of the scalar field to curvature, as specified in (\ref{PLP}). Beyond the slow-roll parameters introduced in this paper, we also provide the key expressions formulated in terms of the Hubble flow parameters as well.

During slow-roll inflation, the following relations apply
\begin{eqnarray}
\label{A1}
&&H=H_{E},~~~\phi=\phi_{E},\\
\label{A2}
&&V(\phi)\simeq\left(\frac{V_{E}(\phi)}{1-n}\right)\left[\frac{V_{E}(\phi)}{V_{E(\ast)}}\right]^{n},\\
\label{A3}
&&F(\phi)\simeq\left(\frac{1}{1-n}\right)\left[\frac{V_{E}(\phi)}{V_{E(\ast)}}\right]^{n},\\
\label{A4}
&&\omega(\phi)\simeq\left[\frac{V_{E}(\phi)}{V_{E(\ast)}}\right]^{n},
\end{eqnarray}
where $V_{E(\ast)}$ is the value of the potential for the case $n=0$ at the crossing of the Hubble radius, and the potential deformation parameter is $-1<n<1$.

The expressions for the slow-roll parameters employed in this work are as follows~\cite{Baumann:2014nda,Chervon:2019sey}
\begin{eqnarray}
\label{A5}
&&\epsilon=\frac{1}{2}\left(\frac{V'_{E}}{V_{E}}\right)^{2},\\
\label{A6}
&&\delta=\frac{V''_{E}}{V_{E}}-\frac{1}{2}\left(\frac{V'_{E}}{V_{E}}\right)^{2}=\eta-\epsilon,\\
\label{A7}
&&\xi=\frac{V'_{E}V'''_{E}}{V^{2}_{E}}-\frac{3}{2}\frac{V''_{E}}{V_{E}}\left(\frac{V'_{E}}{V_{E}}\right)^{2}+
\frac{3}{4}\left(\frac{V'_{E}}{V_{E}}\right)^{4}.
\end{eqnarray}

Consequently, for a known scalar field potential $V = V(\phi)$, the reconstruction of the functional dependence $\delta = \delta(\epsilon)$ is reduced to the determination of the relation $\eta = \eta(\epsilon)$ in explicit form.

The Hubble flow functions are defined as~\cite{Martin:2013tda}
\begin{eqnarray}
\label{A8}
&&\epsilon_{1}=\frac{1}{2}\left(\frac{V'_{E}}{V_{E}}\right)^{2},\\
\label{A9}
&&\epsilon_{2}=2\left[\left(\frac{V'_{E}}{V_{E}}\right)^{2}-\frac{V''_{E}}{V_{E}}\right],\\
\label{A10}
&&\epsilon_{3}=2\left[\frac{V'_{E}V'''_{E}}{V^{2}_{E}}-3\frac{V''_{E}}{V_{E}}\left(\frac{V'_{E}}{V_{E}}\right)^{2}+
2\left(\frac{V'_{E}}{V_{E}}\right)^{4}\right],
\end{eqnarray}

The relations between the slow-roll parameters and the Hubble flow functions are given by expressions
\begin{eqnarray}
\label{A11}
&&\epsilon = \epsilon_{1},~~~~~~~~~~~~~~~~~~~~~~~\epsilon_{1} = \epsilon, \\
\label{A12}
&&\delta = \epsilon_{1} - \frac{1}{2}\epsilon_{2},~~~~~~~~~~~~~~\epsilon_{2} = 2(\epsilon - \delta), \\
\label{A13}
&&\xi = \epsilon_{1}^{2} - \frac{3}{2}\epsilon_{1}\epsilon_{2} + \frac{1}{2}\epsilon_{3},~~~
\epsilon_{3} = 2\xi + 4\epsilon^{2} - 6\epsilon\delta.
\end{eqnarray}

The model-independent condition for the exit from the inflationary stage is given by
\begin{eqnarray}
\label{A14}
&&\epsilon-\delta=\frac{1}{2}\epsilon_{2}>0.
\end{eqnarray}

The difference in the e-folds number between the end of inflation and the Hubble radius crossing can be defined as follows
\begin{equation}
\label{A15}
\Delta N=\frac{1}{2}\int^{\epsilon_{e}}_{\epsilon_{\ast}}\frac{d\epsilon}
{\epsilon(\epsilon-\delta(\epsilon))}=\int^{\epsilon_{1(e)}}_{\epsilon_{1(\ast)}}\frac{d\epsilon_{1}}
{\epsilon_{1}\epsilon_{2}(\epsilon_{1})},
\end{equation}
where $\epsilon_{e},\epsilon_{1(e)}=1$, and $\epsilon_{\ast},\epsilon_{1(\ast)}$ are the values of the first slow-roll parameter and the Hubble flow parameter at the crossing of the Hubble radius.

The scalar field excursion between the end of inflation and the Hubble radius crossing can be written as
\begin{eqnarray}
\label{A16}
\nonumber
&&\frac{|\Delta\phi|}{M_{P}}=\frac{1}{\sqrt{2}}\int^{\epsilon_{e}}_{\epsilon_{\ast}}\frac{d\epsilon}{\sqrt{\epsilon}(\epsilon-\delta(\epsilon))}\\
&&=\sqrt{2}\int^{\epsilon_{1(e)}}_{\epsilon_{1(\ast)}}\frac{d\epsilon_{1}}
{\sqrt{\epsilon_{1}}\epsilon_{2}(\epsilon_{1})}.
\end{eqnarray}

At the Hubble radius crossing ($k=aH$), the scale of inflation $H_{\ast}$ and energy scale of inflation $V^{1/4}_{\ast}$ are defined as follows
\begin{eqnarray}
\label{A17}
&&\frac{H_{\ast}}{M_{P}}=\frac{\pi}{2}\sqrt{2r_{E}A_{S}}=\frac{\pi}{2}\sqrt{\frac{2rA_{S}}{1-n}},\\
\label{A18}
&&\frac{V^{1/4}_{\ast}}{M_{P}}=\left(\frac{3H^{2}_{\ast}}{1-n}\right)^{1/4},
\end{eqnarray}
where $A_{S}=2.1\times10^{-9}$.

Tensor-to-scalar ratio $r$ and the spectral tilt of tensor perturbations are given by expressions
\begin{eqnarray}
\label{A19}
&&r=\frac{A_{T}}{A_{S}}=16\epsilon(1-n)=16\epsilon_{1}(1-n),\\
\label{A20}
&&n_{T}=-\frac{r}{8}=-2\epsilon(1-n)=-2\epsilon_{1}(1-n).
\end{eqnarray}

The spectral index of scalar perturbations and its running are expressed as
\begin{eqnarray}
\label{A21}
&&n_{S}-1=-2(2-n)\epsilon_{\ast}+2\delta_{\ast}\nonumber\\
&&=2(n-1)\epsilon_{1\ast} - \epsilon_{2\ast},\\
\label{A22}
&&\alpha_{S}=(-8 + 4n)\epsilon^{2}_{\ast} + (10 - 4n)\delta_{\ast}\epsilon_{\ast} - 2\xi_{\ast} \nonumber\\
&&=2(n-1)\epsilon_{1\ast}\epsilon_{2\ast} - \epsilon_{3\ast}.
\end{eqnarray}

The deviations from the minimal coupling case in the cosmological perturbation parameters are given by the following expressions
\begin{eqnarray}
\label{A23}
&&r=(1-n)r_{(E)},\\
\label{A24}
&&n_{T}=(1-n)n_{T(E)},\\
\label{A25}
&&\Delta n_{S}=n_{S}-n_{S(E)}=2n\epsilon_{\ast}=2n\epsilon_{1\ast}=\frac{n}{8}r_{(E)},\\
\label{A26}
&&\Delta\alpha_{S}=\alpha_{S}-\alpha_{S(E)}=4n\epsilon_{\ast}(\epsilon_{\ast}-\delta_{\ast})\\
&&=2n\epsilon_{1\ast}\epsilon_{2\ast}=\frac{n}{8}r_{(E)}\left(1-n_{S(E)}-\frac{r_{(E)}}{8}\right).
\end{eqnarray}

When the cosmological perturbation parameters for the case of the Einstein gravity $\{r_{(E)}, n_{T(E)}, n_{S(E)}, \alpha_{S(E)}\}$ are known, expressions (\ref{A23})--(\ref{A26}) are sufficient to evaluate the corresponding values $\{r, n_{T}, n_{S},\alpha_{S}\}$ in the presence of the non-minimal coupling of the scalar field to curvature.
The values of the cosmological perturbation parameters for Einstein gravity are given for a wide range of inflationary models in~\cite{Martin:2013tda}.

\bibliography{ref_short}

\end{document}